\documentclass[fleqn,usenatbib]{mnras}

\usepackage{newtxtext,newtxmath}
\usepackage{color}
\usepackage[T1]{fontenc}
\usepackage{ae,aecompl}
\usepackage{xcolor}
\usepackage{ulem}
\usepackage{graphicx}	% Including figure files
\usepackage{amsmath}	% Advanced maths commands
\usepackage{cases}
\usepackage{array,multirow}
\newcommand{\Msol}{\;M_{\odot}}

\newcommand{\pbhm}{M_{\rm BH,1}}
\newcommand{\sbhm}{M_{\rm BH,2}}
\newcommand{\mstar}{\;M_{\star}}
\newcommand{\UpTDE}{\Upsilon_{\rm TDE}}
\newcommand{\UpDC}{\Upsilon_{\rm DC}}

\title[TDEs by SMBHBs]{Tidal Disruption Events by Compact Supermassive Black Hole Binaries}

\author[T.Ryu et al.]{Taeho Ryu$^{1}$\thanks{E-mail:tryu@mpa-garching.mpg.de} , Alessandro A. Trani$^{2,3}$, Nathan W. C. Leigh$^{4,5}$\\
$^{1}$Max Planck Institute for Astrophysics, Karl-Schwarzschild-Strasse 1, 85748 Garching, Germany\\
$^{2}$Department of Earth Science and Astronomy, College of Arts and Sciences, The University of Tokyo, 3-8-1 Komaba, Meguro-ku, Tokyo 153-8902, Japan \\
$^{3}$Okinawa Institute of Science and Technology, 1919-1 Tancha, Onna-son, Okinawa 904-0495, Japan \\
$^{4}$Departamento de Astronom\'ia, Facultad Ciencias F\'isicas y Matem\'aticas, Universidad de Concepci\'on, Av. Esteban Iturra s/n Barrio Universitario,\\ Casilla 160-C, Concepci\'on, Chile \\
$^{5}$Department of Astrophysics, American Museum of Natural History, New York, NY 10024, USA \\
}

\date{Accepted XXX. Received YYY; in original form ZZZ}

\pubyear{2022}

\begin{document}
\label{firstpage}
\pagerange{\pageref{firstpage}--\pageref{lastpage}}
\maketitle

\begin{abstract}
Stars can be tidally destroyed or swallowed by supermassive black hole binaries (SMBHBs). Using a large number of few-body simulations, we investigate the enhancement and suppression of full and partial disruption and direct capture events by hard SMBHBs with wide ranges of key parameters, i.e., the primary BH mass ($M_{\rm BH, 1}= 10^{5}-10^{8}M_{\odot}$), the binary mass ratio ($10^{-3}-1$), the ratio of the binary semimajor axis to the hardening radius ($10^{-4}-1$), the binary eccentricity ($0.0-0.9$) and the stellar mass ($0.3-3M_{\odot}$). This is a significant extension of the parameter space compared to previous work. We show that the encounter probabilities of all three events are well-described by the encounter cross section. The probability of full tidal disruption events (FTDEs) by SMBHBs can be enhanced by up to a factor of $40-50$ or suppressed by up to a factor of $10$, relative to that by single BHs, depending on the binary parameters. Relativistic effects can provide an additional enhancement of the FTDE probability by less than a factor of $2-3$ for $M_{\rm BH, 1}>10^{7}M_{\odot}$. We provide a fitting formula for the FTDE probability by SMBHBs that works for a wide range of parameters. Partial disruption events can occur multiple times before FTDEs or direct captures, and their probabilities can be greater than that of FTDEs by a factor of three. Because partial disruption events can induce stellar spins and mass loss, and change the orbits, it can significantly affect the overall FTDE rate and the shape of the light curves.
\end{abstract}


%%%%%%%%%%%%%%%%%%%%%%%%%%%%%%%
\section{Introduction}

It is believed that almost every galaxy harbors a supermassive black hole (SMBH) at the center \citep{KormendyHo2013}. Occasionally many gravitational encounters place some stars on nearly parabolic orbits around the SMBH. When the pericenter of the orbit is sufficiently small, the stars are disrupted by the tidal force of the SMBH at the first pericenter passage \citep{hills88}.  These events are called tidal disruption events (TDEs). Roughly, half the debris is bound and the other half is unbound. Ultimately, the bound debris returns to the SMBH. The energy liberated in this process can generate a bright flare. Around 100 candidates have been detected, and the number will increase rapidly with detections by ongoing \citep[e.g., eROSITA][]{Sazonov+2021} and future surveys \citep[e.g.,  LSST][]{BricmanGomboc2020}

SMBH binaries (SMBHBs) at the galactic centers, which are a natural byproduct of galaxy mergers, can disrupt stars as well \citep[e.g.][]{merritt13,li15}. In fact, some observed TDEs have been suggested as the events by SMBHBs \citep[e.g.,][]{Liu+2014,Huang+2021}. The evolution of two SMBHs up to their mergers in merging galaxies  can be generally described by three stages, depending on different mechanisms for shrinking the binary orbit. Initially, two SMBHs sink towards the galactic center by dynamical friction. When they become close, they finally form a binary. Then, their orbits shrink via interactions with surrounding stars. As binaries become hard, a significant fraction of stars are ejected, resulting in empty phase space regions around the binary (``empty loss cone regime''). The distance at which a SMBH binary with the primary mass $M_{\rm BH,1}$ and the secondary mass $M_{\rm BH,2}$ becomes hard is roughly \citep{Quinlan1996},
\begin{align}\label{eq:rh}
    r_{\rm h} = \frac{G M_{\rm BH,1}M_{\rm BH,2}}{4(M_{\rm BH,1}+M_{\rm BH,2})\sigma^{2}},
\end{align}
where $\sigma$ is the the velocity dispersion of surrounding stars. The subscript 1 refers to the primary BH and 2 to the secondary BH hereafter. Finally, at sufficiently small orbital separations, gravitational wave emission drives the SMBHBs to merge.

Tidal disruption events can be caused by SMBHBs at any of the aforementioned evolutionary stages \citep{Li+2017}. See \citet{CoughlinCoughlinLodato} for a recent review of TDEs by SMBHBs. This physics problem has been studied using purely three-body scattering experiments \citep[e.g.,][]{Ivanov+2005,Chen+2009,Chen+2011,Darbha+2018} or by combining scattering experiments and hydrodynamics simulations \citep{Coughlin+2017,CoughlinArmitage}. In general, until the separation of the two BHs becomes comparable to the influence radius of the primary BH, the TDE rate is not greatly enhanced by the secondary BH \citep{Li+2017}. As the two distances become comparable, the secondary can affect the stellar orbits inside the influence radius, leading to possible significant increase in the TDE rate \citep{Ivanov+2005} via Kozai-Lidov mechanism \citep{Kozai1962,Lidov1962}. When the binary becomes hard, chaotic interactions, in addition to the Kozai-Lidov mechanism, induced by the secondary BH contribute to the enhancement of the rate \citep{reinoso22}. The decaying binaries can result in a short-lived ($\lesssim 1$ Myr) burst of the events at a rate that can be a few orders of magnitude greater than that for single SMBHs \citep{Chen+2009, Chen+2011}. When the binaries have ejected most of the surrounding stars, and stars at large distances approach the binary on nearly radial orbits, the impact of the binaries on the TDE rate appears to be relatively modest. In fact, it has been suggested that the event rate can be enhanced by a factor of a few \citep{Coughlin+2017, Darbha+2018} or even suppressed by one order of magnitude \citep{Chen+2008}, depending on the adopted assumption of loss-cone re-population. 

In this work, using highly accurate three-body scattering experiments, we investigate full, partial disruption events and direct captures (see \S\ref{sub:outcome}) of main-sequence stars by hard SMBHBs over an extensively large range of parameter space of the key parameters, i.e., the primary BH mass $\pbhm$, secondary BH mass $\sbhm$, the semimajor axis of the SMBHB $a$, eccentricity of the SMBHB $e$ and stellar mass $M_{\star}$. This is a significant extension of the parameter range covered by previous works. In particular, by considering a wide range of black hole masses, we study whether full disruptions are enhanced or suppressed in those cases where stars are fully disrupted only by the secondary black hole. We also examine the impact of the post-Newtonian (PN) terms on the probabilities of the events. We provide a fitting formula for the relative probability of full disruptions by SMBHBs relative to those by single SMBHs that work for the parameter space that we consider. We also study the frequency of partial disruptions and discuss their potentially important implications.

This paper is organized as follows. In \S\ref{Sec:setup}, we provide a detailed description of our simulation setup, including the initial conditions (\S\ref{subsec:initialcondition}) and simulation termination criteria (\S\ref{sub:outcome}). In \S\ref{sec:result}, we present the results of our simulations using the concept of the cross-section (\S\ref{sec:crosssection}). Then, we discuss the implications of partial disruptions and a few caveats and future improvement in \S\ref{sec:discussion} before concluding and summarizing this paper in \S\ref{sec:concluson}.

\section{Numerical setup}\label{Sec:setup}

We perform a large number of three-body scattering experiments between a relatively compact SMBHB and a main-sequence star using {\texttt {TSUNAMI}} \citep{Trani+2019}. {\texttt {TSUNAMI}} is a very accurate and fast few-body code that is suitable for studying the evolution of few-body systems in which there are frequent gravitational encounters at short distances between masses with large contrasts. This code has been used to study a variety of astrophysical problems \citep[e.g.][]{Trani+2019b, trani2020,trani2020b,trani2022b}, showing the code's accuracy and precision. The {\texttt {TSUNAMI}} code employs several well-established techniques to achieve this. First, we solve the equations of motions derived from a logarithmic Hamiltionian \citep{mik99a}. This makes the integration timestep practically independent of the eccentricities of close binaries, avoiding the integration to halt during close encounters. Second, we improve the accuracy of the integration with the use of Bulirsch-Stoer extrapolation \citep{stoer1980}. Finally, to reduce round-off errors, we use a chain-coordinate system for the particles' positions and velocities, rather than the commonly used center-of-mass coordinates \citep{mik90}. {\texttt {TSUNAMI}} includes also Post-Newtonian corrections responsible for the pericenter shift (1PN and 2PN) and the quadrupole gravitational radiation (2.5PN).

\begin{table}
    \centering
	\begin{tabular}{c c c}
		\hline
		Parameter & Range & Selected values \\
		$\pbhm~[\Msol]$  &  $10^{5}- 10^{8}$ & $10^{5}$, $10^{6}$, $10^{7}$, $3\times10^{7}$, $10^{8}$\\ 
		$q$   & $10^{-3}-1$ & $10^{-3}$, $10^{-2}$, $10^{-1}$, $1$ \\
		$a/r_{\rm h}$   & $10^{-4}-1$ & $10^{-4}$, $10^{-3}$, $10^{-2}$, $10^{-1}$, $1$  \\
		$e$   & $0-0.9$ & 0, 0.1, 0.3, 0.4, 0.5, 0.7, 0.9 \\
		$\mstar~[\Msol]$   & $0.3-3$ & 0.3, 1, 3 \\
    \hline
	\end{tabular}
		\caption{The ranges and selected values of the key parameters considered in the experiments, (from top to bottom) $\pbhm$: the primary black hole mass, $q=\sbhm/\pbhm$: mass ratio, $a/r_{\rm h}$: the ratio of the semimajor axis to the hardening radius (Equation~\ref{eq:rh}), $e$: the eccentricity of the SMBHB, $\Msol$: the stellar mass. We impose two constraints on the range, 1) $a> a_{\rm GW}$ where $a_{\rm GW}$ is the semimajor axis at which the merger time due to gravitational radiation emissions is greater than $10^{4}\times$ (binary period) and 2) $\sbhm\geq 10^{3}$. } \label{tab:testlist}
\end{table}

\subsection{Initial conditions}\label{subsec:initialcondition}

We consider SMBHBs with the primary black hole mass of $10^5\Msol\leq\pbhm\leq 10^{8}\Msol$ and the mass ratio of $10^{-3}\leq q=\pbhm/\sbhm \leq 1$, with a constraint of $\sbhm\geq10^{3}\Msol$. The binary's semimajor axis is $10^{-4}\leq a/r_{\rm h}\leq 1$. To calculate $r_{\rm h}$, we compute $\sigma$ using the $M_{\rm BH}-\sigma$ relation by \citet{KormendyHo2013} with the total black hole mass $\pbhm+\sbhm$. For a very short semimajor, the gravitational wave emission can shrink the binary to its merger in a time scale less than the maximum integration time of our simulations, i.e., $10^{4}P$ where $P$ is the binary period. So we only consider the semimajor axis larger than the distance $a_{\rm GW}$ at which the binary lifetime determined by the gravitational wave emission is longer than the integration time. The range of the binary eccentricity is \textit{$0.0\leq e\leq 0.9$}. We uniformly sample the argument of pericenter and the mean anomaly between 0 and $2\uppi$. The binary's orbital plane is always in the x-y plane of the coordinate system.

The star initially approaches the center of the mass of the SMBHB on a parabolic orbit ($1-e\simeq 10^{-9}$). We uniformly sample the pericenter distance of the incoming stellar orbit within the range of [0, $2a$]. The initial distance of the star from the SMBHB's center of mass is $100a$. We consider three different stellar masses, $\mstar=0.3,~ 1$ and $3\Msol$. The inclination of the parabolic orbit relative to the binary's orbital plane, the argument of pericenter, and the longitude of the ascending node are uniformly sampled. 

The range and selected values of the key parameters considered in our simulations are summarized in Table~\ref{tab:testlist}.

\subsection{Outcome and termination criteria}\label{sub:outcome}

The possible outcomes in these experiments are 1) ejection, 2) full tidal disruption event (FTDE), 3) direct capture (DC) and 4) triple formation (bound orbit). The stars can be destroyed by either FTDEs or DCs. A FTDE occurs when the pericenter distance of the star's orbit is shorter than the tidal radius of one of the BHs. The stars can be swallowed whole when the pericenter distance is even shorter than the radius of one of the BHs. On the other hand, the stars can escape from the SMBHB's potential to infinity or form a triple with the binary. Note that the triple formation outcome is not a definitive one, as the triple might not be stable at all times, and eventually end up in an ejection or a tidal disruption. Nonetheless, we choose a relatively long total integration time, to ensure that unstable triples may evolve into other outcomes. 

Our termination conditions are determined to properly identify the four different outcomes. We terminate the simulations when one of the following conditions are satisfied, and classify the outcome accordingly.

\begin{enumerate}
    \item \textit{Ejection}: When the star has left the sphere of radius $100 a$  with a positive specific orbital energy relative to the potential of the SMBHB. 
    
    \item \textit{Full disruption}: When the pericenter distance $r_{\rm p}$ is smaller than the tidal radius $R_{\rm TDE}$ (see \S\ref{sub:rt}) of one of the BHs. For this study, we take  the numerically estimated tidal radius $\mathcal{R}_{\rm t}$ for realistic main-sequence stars using relativistic hydrodynamics simulations in \citet{Ryu+2020a} (see \S\ref{sub:rt}). 

    \item \textit{Direct capture}: When the pericenter distance is smaller than the characteristic pericenter distance $\mathcal{R}_{\rm DC}$ of one of the BHs. Here, $\mathcal{R}_{\rm DC}$ is the distance below the star is swallowed whole. For parabolic orbits, $\mathcal{R}_{\rm DC} = 2~r_{\rm Sch}$, where $r_{\rm Sch}$ is the Schwarzschild radius. But we do not make any assumptions on the stellar orbits when DCs occur. So we take a conservative choice for $\mathcal{R}_{\rm DC}$, namely, $r_{\rm Sch}$. 
    
    \item \textit{Bound}: When the integration time exceeds $10^{4}~P$ where $P$ is the orbital period of the SMBHB without the star being ejected, tidally disrupted or directly captured.
    
\end{enumerate}

To correctly identify FTDEs and DCs, we first flag the star as a TDE candidate when the distance of the star from any BH is smaller than $R_{\rm TDE}$. If the distance becomes shorter than $\mathcal{R}_{\rm DC}$, the event is classified as a DC and we stop the simulation. If the distance never becomes smaller than $\mathcal{R}_{\rm DC}$ and starts to increase, the event is classified as a TDE and we stop the simulation.

In this paper, we choose specific values of $\mathcal{R}_{\rm TDE}$ and $\mathcal{R}_{\rm DC}$ to quantify the enhancement and suppression factor of TDEs and DCs by SMBHBs, which is defined in terms of the relative cross-sections ($\UpTDE$, defined in Equation~\ref{eq:rel_proba} and $\UpDC$, defined in Equation~\ref{eq:rel_probaDC}). However, as shown in Section~\ref{sec:result}, because the event probability is proportional to the cross section, the values of $\Upsilon$ can be easily adjusted for different choices of $\mathcal{R}_{\rm TDE}$ and $\mathcal{R}_{\rm DC}$ by proper renormalization.

For each set, we perform more than $10^{6}- 10^{7}$ scattering experiments with the termination conditions until we have converging results.

\begin{figure}
	\centering
	\includegraphics[width=8.6cm]{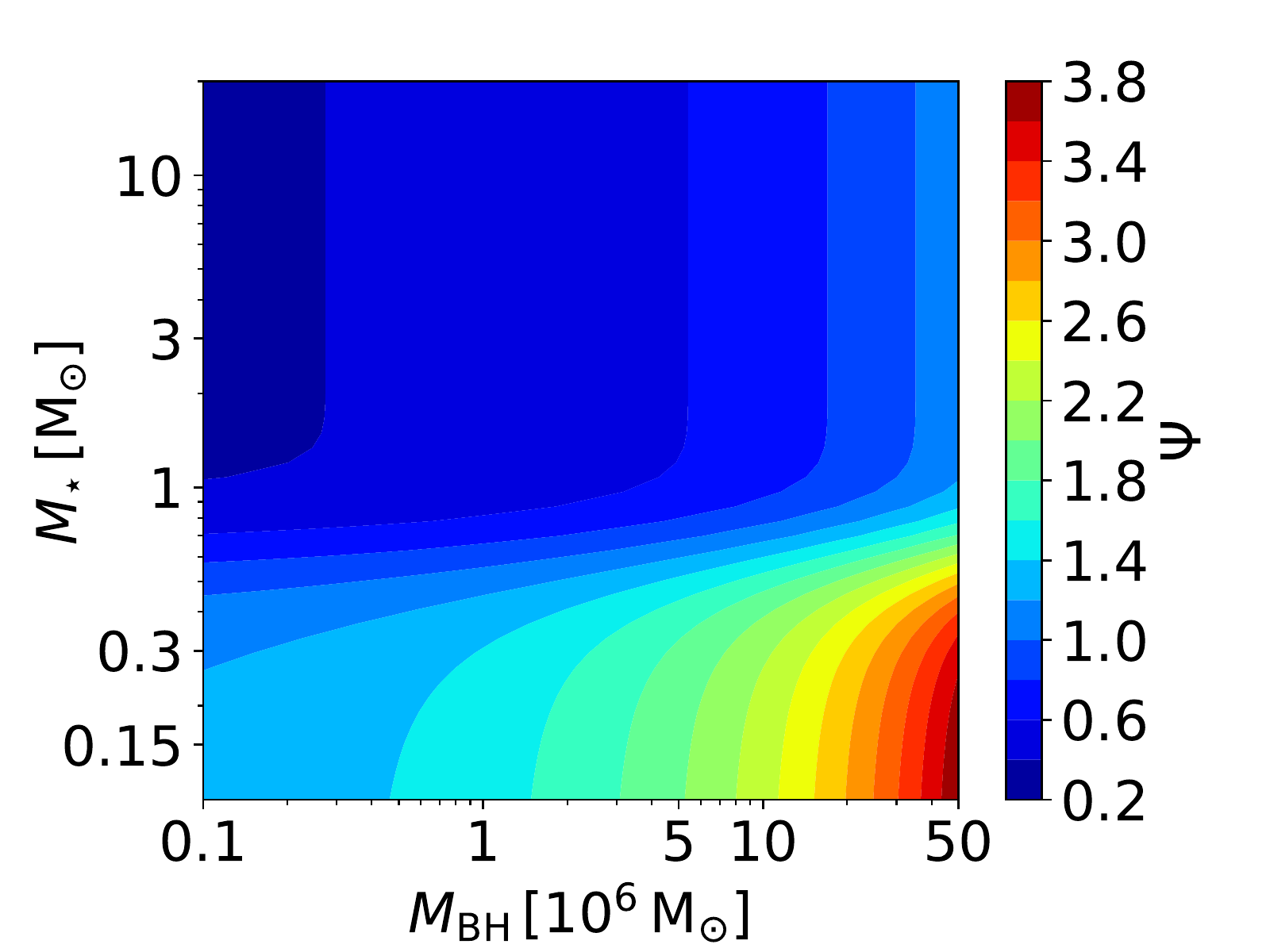}
\caption{The correction factor $\Psi$ of the maximum pericenter distance for full disruptions $\mathcal{R}_{\rm t}$ relative to the canonical tidal radius $r_{\rm t}$, i.e., $\Psi =\mathcal{R}_{\rm t}/ r_{\rm t}$, estimated by taking into account realistic stellar structure and relativistic effects in \citet{Ryu+2020a}. This plot is made using their fitting formulae (Equations~\ref{eq:fit_psistar} and \ref{eq:fit_psibh}). $\Psi$ increases as $M_{\star}$ decreases (due to higher central concentration) and $M_{\rm BH}$ increases (due to more destructive tidal force). }
	\label{fig:psi}
\end{figure}

\begin{figure*}
	\centering
	\includegraphics[width=8.6cm]{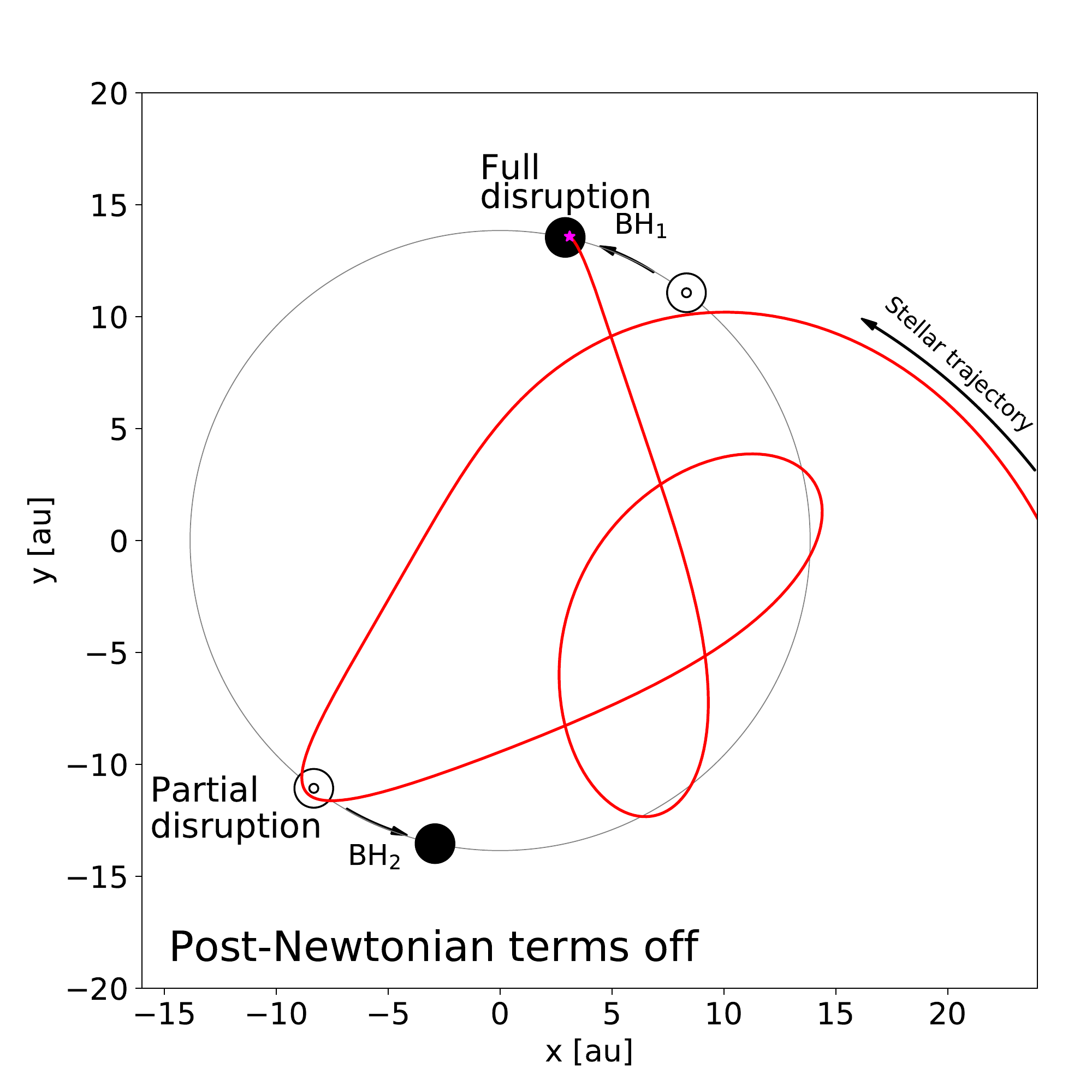}
	\includegraphics[width=8.6cm]{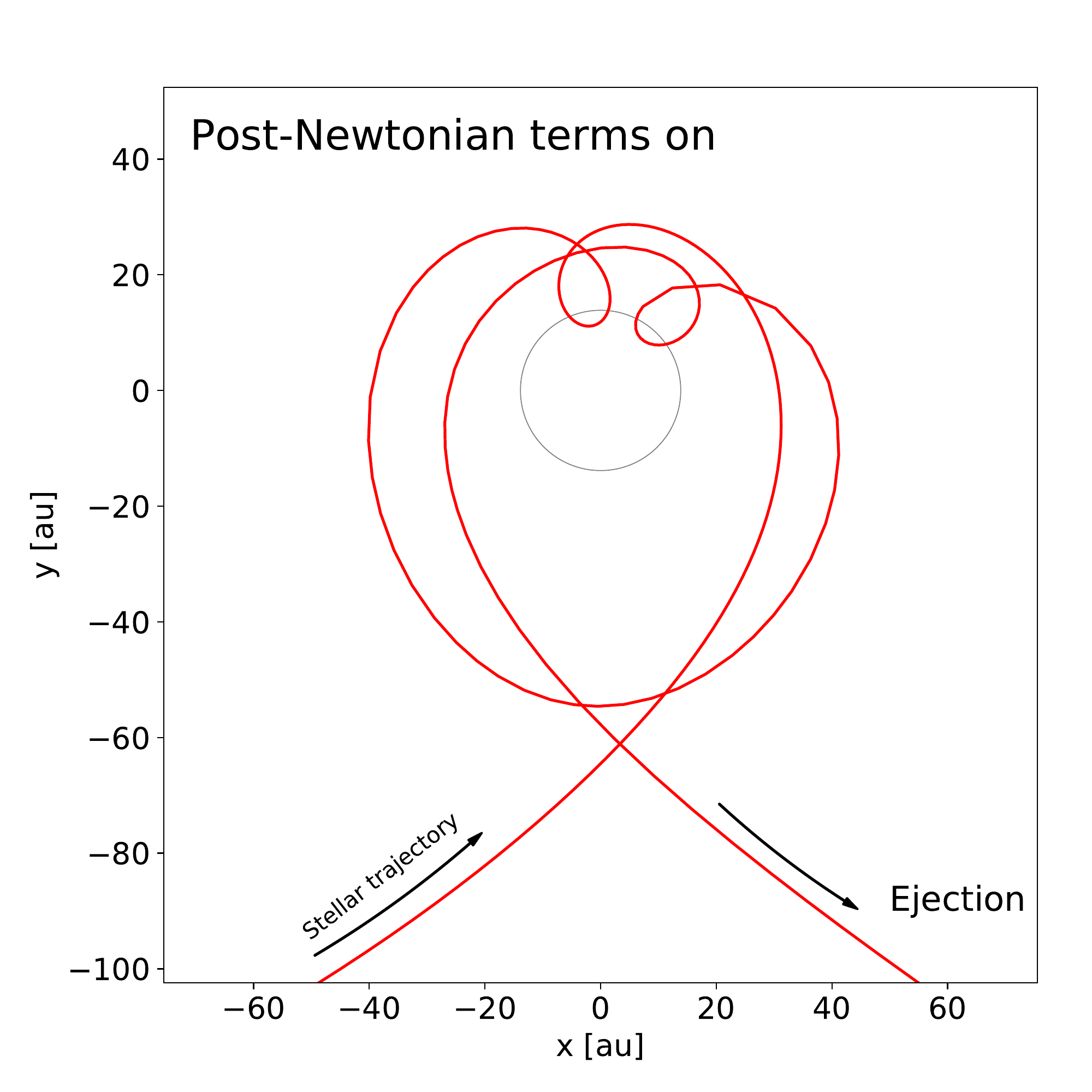}\\
	\includegraphics[width=8.6cm]{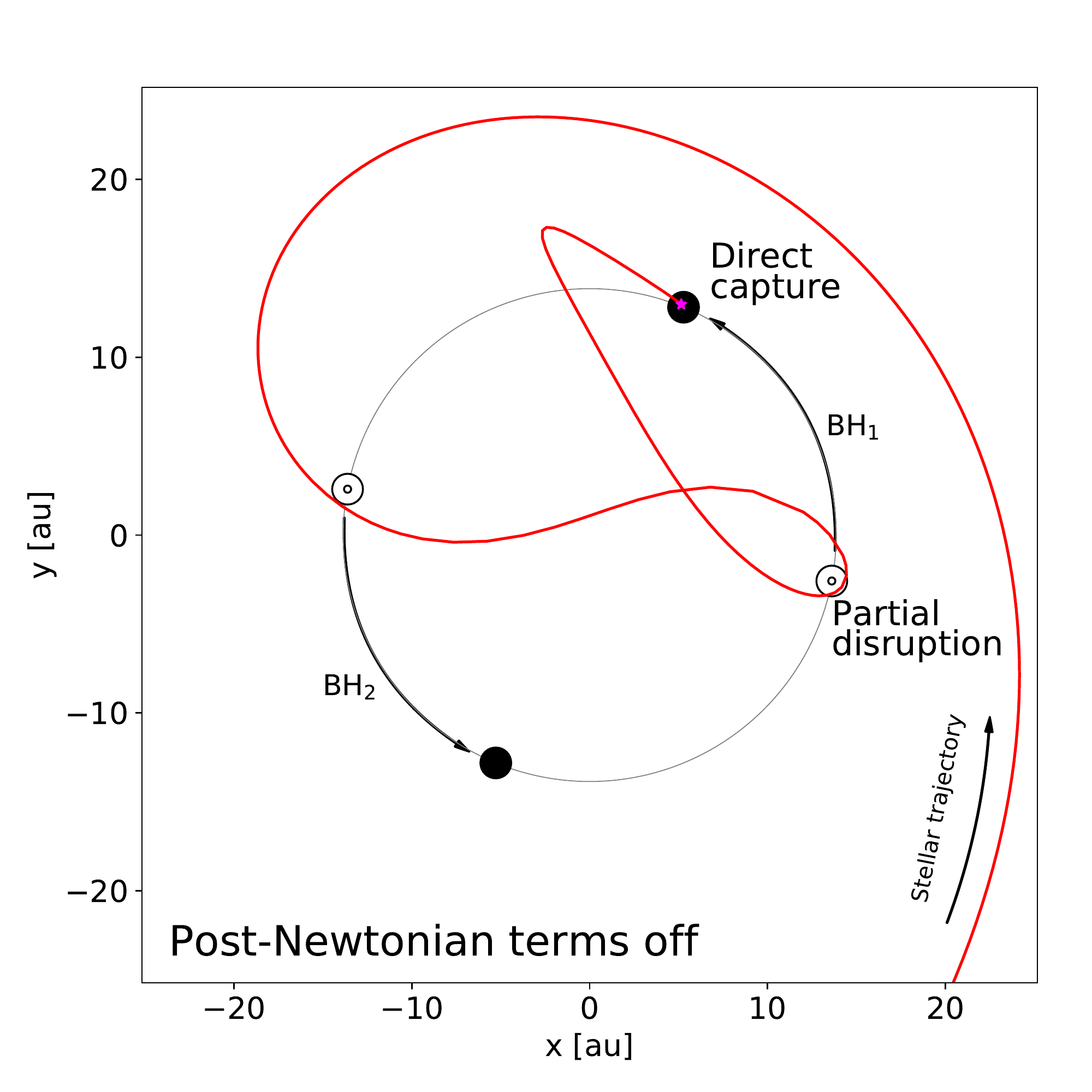}
	\includegraphics[width=8.6cm]{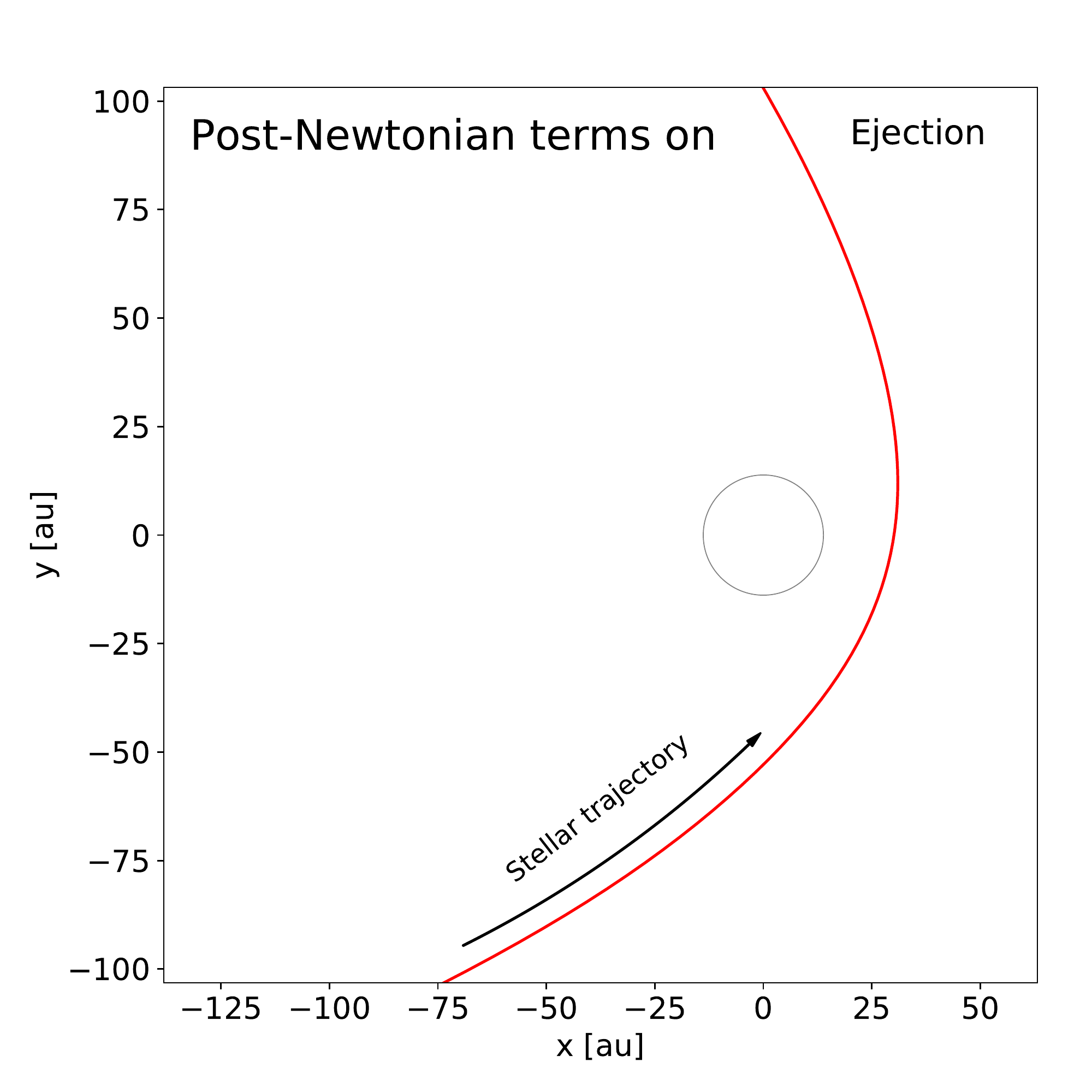}
	\caption{Projected stellar trajectories in the xy-plane around an equal-mass black hole binary with $\pbhm=\sbhm=10^7\Msol$ and $a/r_{\rm h}=10^{-2}$.
	The red lines indicate the trajectories of the star and the large circle around the center is the trajectory of the binary. When the post-Newtonian terms are turned off (the \textit{left} panels), the star ends up being fully disrupted (\textit{top}) or directly captured (\textit{bottom}). The filled dots are the final locations of the BHs and the open dots are the locations of the BHs when a partial disruption occurs. The outer (inner) boundary of the open dots demarcates the maximum distance that yields a partial (full) disruption.
	The \textit{right} panels depict the stellar trajectories in the simulations starting with the same initial conditions as the runs in the \textit{left} panels when the post-Newtonian terms are turned on. The final outcomes are ejections. }
	\label{fig:ex_traj}
\end{figure*}

\subsection{Tidal radius}\label{sub:rt}
We adopt the correction factor $\Psi$ for the maximum pericenter distance $\mathcal{R}_{\rm t}$ yielding full disruptions by \citet{Ryu+2020a} to its order-or-magnitude estimate $r_{\rm t}$,
\begin{align}
\label{eq:tidalradius}
r_{\rm t}&=\left(\frac{M_{\rm BH}}{M_{\star}}\right)^{1/3}R_{\star},
\end{align}
where $R_{\star}$ is the radius of the star. So $\mathcal{R}_{\rm t}=\Psi r_{\rm t}$. To measure the correction factor, they performed a suite of general relativistic hydrodynamics simulations for tidal disruption events of realistic intermediate-age main-sequence stars at the first pericenter passage. They find that the correction factor is a function of $M_{\star}$ and $M_{\rm BH}$, which are separable. So the correction factor $\Psi$ can be expressed as the product of a $M_{\star}$-dependent term and a $M_{\rm BH}$-dependent term,
\begin{align}
    \Psi(M_{\rm BH},M_{\star})= \Psi_{\rm BH}(M_{\rm BH})\Psi_{\star} (M_{\star}),
\end{align}
where 
\begin{align}
\Psi_{\star}(M_{\star})& = \frac{1.47+ ~\exp[(M_{\star}/\Msol-0.669 )/0.137]}{1 + 2.34~\exp[(M_{\star}/\Msol-0.669)/0.137]},\label{eq:fit_psistar}\\
\Psi_{\rm BH}(M_{\rm BH})&=0.80 + 0.26~\left(\frac{M_{\rm BH}}{10^{6}\Msol}\right)^{0.5}.\label{eq:fit_psibh}
\end{align}
$\Psi_{\star}$ incorporates the change in the tidal radius associated with stellar internal structure. Figure~\ref{fig:psi} depicts the correction factor as a function of $M_{\star}$ and $M_{\rm BH}$. As shown in the figure, for a fixed $M_{\rm BH}$, $\Psi_{\star}$ decreases as $M_{\star}$ increases because more massive stars are more centrally concentrated so that they can get closer to the BH until they are fully disrupted. Remarkably, because of the opposite sense of $\Psi_{\star}$ and $r_{\rm t}$ with respect to $M_{\star}$, $\mathcal{R}_{\rm t}$, measured in physical units, is nearly independent of $M_{\star}$ for a given $M_{\rm BH}$ (see Figure~3 in \citealt{Ryu+2020a}). On the other hand, $\Psi_{\rm BH}$ is associated with relativistic effects. For a fixed $M_{\star}$, $\Psi_{\rm BH}$ grows with $M_{\rm BH}$ because of more destructive relativistic tidal stress for higher $M_{\rm BH}$.

For each set, we run another simulation with the post-Newtonian terms switched on. For the Newtonian case with the post-Newtonian terms ``off'', we only include the correction associated with stellar structure, 
\begin{align}\label{eq:Rt}
    \mathcal{R}_{\rm t} = \Psi_{\star}(M_{\star})~ r_{\rm t}.
\end{align}
On the other hand, for the relativistic case with the post-Newtonian terms ``on'', we include the whole correction,
\begin{align}\label{eq:RtBH}
    \mathcal{R}_{\rm t} = \Psi(M_{\star}, M_{\rm BH})~ r_{\rm t}.
\end{align}

$\mathcal{R}_{\rm t}/\mathcal{R}_{\rm DC}>1$ for sufficiently small $M_{\rm BH}$ (e.g., $\lesssim 10^{7}\Msol$). However because $\mathcal{R}_{\rm t}/\mathcal{R}_{\rm DC}$ decreases with $M_{\rm BH}$, $\mathcal{R}_{\rm t}$ can become smaller than $\mathcal{R}_{\rm DC}$ for sufficiently large $M_{\rm BH}$ ($\gtrsim 5\times 10^{7}\Msol$). Here, we define the maximum black hole mass $\widehat{M}_{\rm BH}$ capable of disrupting stars, or the mass at which $\mathcal{R}_{\rm t} = \mathcal{R}_{\rm DC}$. So $\mathcal{R}_{\rm t} - \mathcal{R}_{\rm DC}>0$ is equivalent to $M_{\rm BH}<\widehat{M}_{\rm BH}$.

We note that \citealt{Ryu+2020a} considers the black hole mass $10^{5}\leq M_{\rm BH}/M_{\odot}\leq 5\times 10^{7}$ to find the fitting formula (Equations~\ref{eq:fit_psistar} and \ref{eq:fit_psibh}). Therefore, the correction factor for $\pbhm=10^{8}\Msol$, the largest mass considered in our simulations, may not be as accurate as that for lower $\pbhm$. Nonetheless, we consider such a large $\pbhm$ to investigate the role of the secondary BH in the case where the primary BH is not capable of disrupting stars. As will be shown later, the enhancement or suppression of FTDEs by SMBHBs is well described by the encounter cross section (\S\ref{sec:crosssection}), which can be generalized to any choice of $\mathcal{R}_{\rm t}$.

To check if the probability $\UpTDE$ (Equation~\ref{eq:rel_proba}) and $\UpDC$ (Equation~\ref{eq:rel_probaDC}) of FTDEs and DCs by SMBHBs relative to those by single SMBHs , which are our main results, would be affected by a different choice of $R_{\rm TDE}$, we performed extra simulations with $R_{\rm TDE}=r_{\rm t}$. We find that those probabilities are in good agreement between the two cases. This is somewhat expected because, as we will show later, the number of FTDEs and DCs by SMBHBs ($N_{\rm TDE}$ and $N_{\rm DC}$) is proportional to the cross section $\propto R_{\rm TDE}$ and $R_{\rm DC}$, respectively.

\subsection{Examples}\label{sub:example}

The \textit{left} panels of Figure~\ref{fig:ex_traj} show the projected trajectories in the xy-plane of the orbits for stars that end up being fully disrupted (\textit{top} panel) and directly captured (\textit{bottom} panel) by an equal-mass binary with $\pbhm=\sbhm=10^7\Msol$ ($a/r_{\rm h}=10^{-2}$) when the PN terms are turned off. Interestingly, in both cases, the stars have passed the partial disruption zone of one BH (the area between smaller and larger circles) before they are fully disrupted or swallowed whole. This means that, before the stars are destroyed, they would lose some fraction of their masses and acquire spins. Remarkably, as will be shown in \S\ref{sec:ptde}, such close encounters occur prior to a full disruption or direct capture more than once in many cases. We will discuss its implications in \S\ref{subsec:disc_ptde}. For comparison, the \textit{right} panels show the trajectories of the stars that are finally ejected when the PN terms are turned on while the initial conditions are the same.

\begin{figure}
	\centering
	\includegraphics[width=8.6cm]{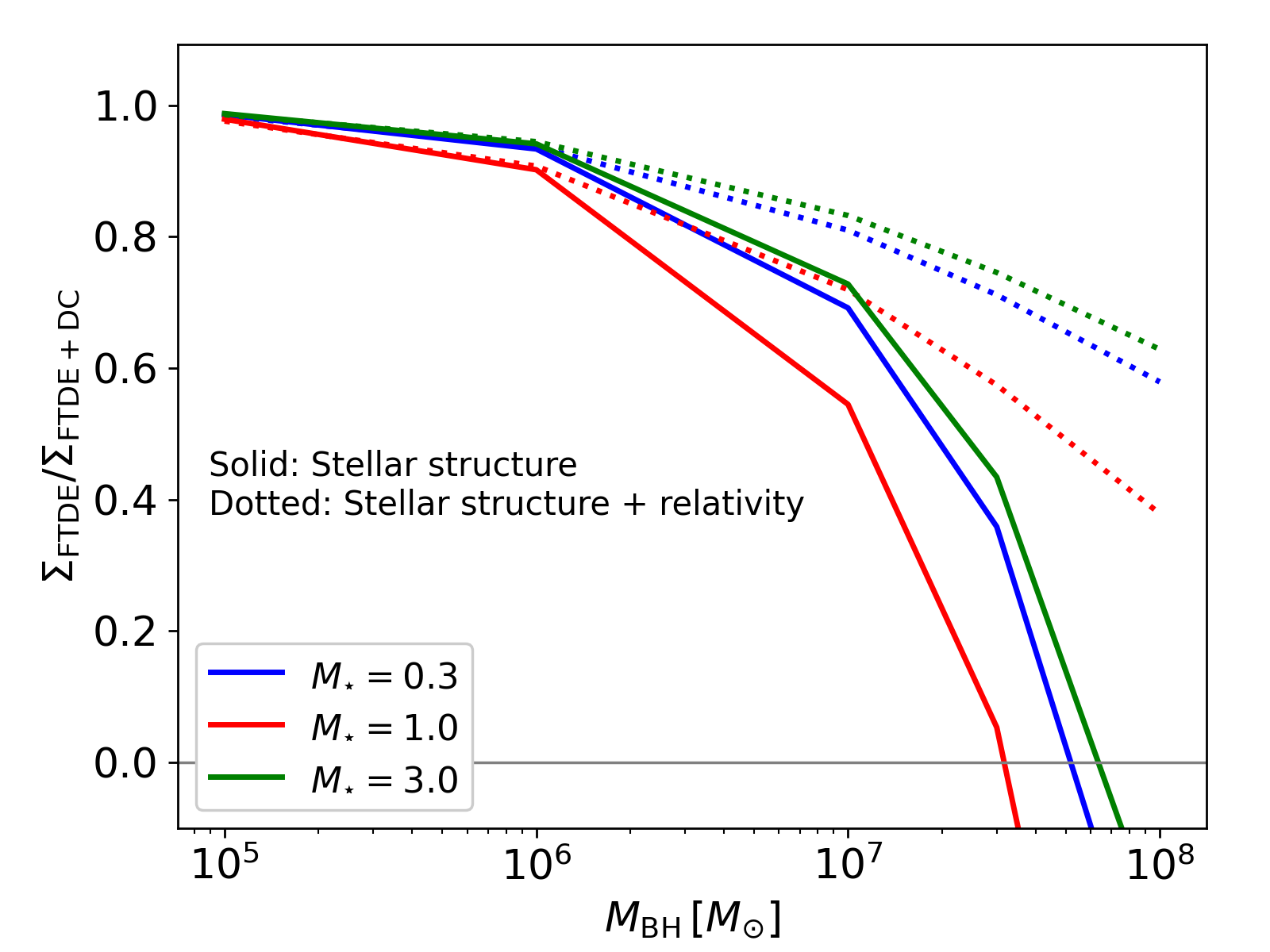}
\caption{The ratio of the cross section for full disruptions $\Sigma_{\rm FTDE}$ to that for star-removing events (full disruptions and direct captures) $\Sigma_{\rm FTDE+DC}$ as a function of the black hole mass $M_{\rm BH}$ for three different stellar masses. The solid (dotted) lines indicate the case where the tidal radius is corrected by realistic stellar structure (realistic stellar structure and relativity) from \citet{Ryu+2020a}. Negative means the cross section for direct captures is greater than that for full disruptions, therefore no disruption events.  }
	\label{fig:cross}
\end{figure}

\begin{figure}
	\centering
	\includegraphics[width=8.6cm]{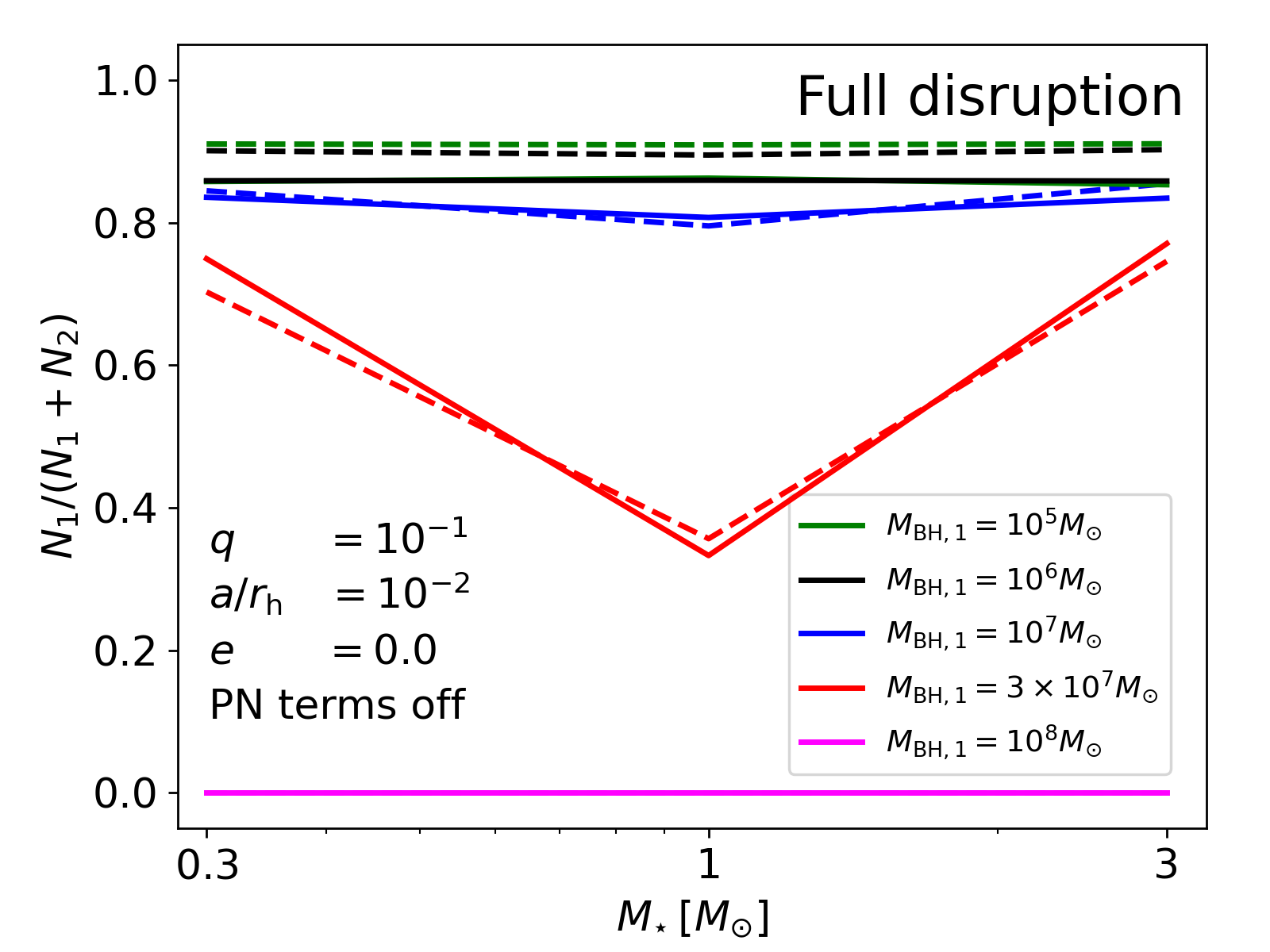}
\caption{The ratio of the number of events $N_{\rm 1}$ for FTDEs by the primary BH in circular unequal-mass SMBHBs with $a/r_{\rm h}=10^{-2}$  to the total number of the events ($N_{1}+N_{2}$), as a function of $M_{\star}$, for different values of $\pbhm$. The dashed lines indicate the values of $\Sigma_{\mathrm{TDE},1}/(\Sigma_{\mathrm{TDE},1}+\Sigma_{\mathrm{TDE},2})$ + 0.2 for a given $\pbhm\leq10^{7}\Msol$ and $\sbhm$. Note that $N_{1}=0$ for $\pbhm=10^{8}$ as expected from $\Sigma_{\mathrm{TDE},1}<0$. }
	\label{fig:eventfraction_mstar}
\end{figure}

\section{Cross sections for full disruptions and direct captures}\label{sec:crosssection}

One of the important concepts to understand the probability of FTDEs and DCs by SMBHBs is the encounter cross section. 
The cross section $\Sigma$ of a strong stellar encounter is proportional to the pericenter distance \footnote{The cross section with gravitational focusing is $\Sigma= \uppi r_{\rm p}^{2} ( 1+ v^{2}/\sigma^{2})$ where $v$ is the escape speed from the point of contact, $\propto 1/r_{\rm p}^{0.5}$. Because $v\gg \sigma$, $\Sigma\propto r_{\rm p}^{2} v^{2}\propto r_{\rm p}$.}. The cross section for FTDEs is $\Sigma_{\rm FTDE}\propto \mathcal{R}_{\rm t} - \mathcal{R}_{\rm DC}$ and that for DC is $\Sigma_{\rm DC}\propto \mathcal{R}_{\rm DC}$. In fact, this cross section was already used in previous work \citep{Chen+2008,Chen+2009,Chen+2011} to describe the FTDE probability by SMBHBs.

As explained in \S\ref{sub:rt}, $\Sigma_{\rm FTDE}$ becomes small as $M_{\rm BH}$ increases and finally zero at $M_{\rm BH}=\widehat{M}_{\rm BH}$, as shown in Figure~\ref{fig:cross}. However, the value of $\widehat{M}_{\rm BH}$ is different for different values of $M_{\star}$.  Because of this BH mass constraint, we consider three different cases for FTDEs by SMBHBs. 
\begin{enumerate}
    \item $M_{\rm BH,1}<\widehat{M}_{\rm BH,1}$: both BHs can disrupt the stars.  The more massive BH has more chances of disrupting or swallowing the stars because of the larger cross section ($[\Sigma_{\rm FTDE,2}+\Sigma_{\rm DC,2}]/[\Sigma_{\rm FTDE, 1} + \Sigma_{\rm DC,1}]\simeq q^{1/3}$). For $\pbhm\ll \widehat{M}_{\rm BH,1}$ (or $\mathcal{R}_{\rm t,1}\gg \mathcal{R}_{\rm DC,1}$), FTDEs would dominate DCs because of its large cross section, $\Sigma_{\rm TDE,1}/(\Sigma_{\rm TDE,1}+\Sigma_{\rm DC,1})=(\mathcal{R}_{\rm t,1}-\mathcal{R}_{\rm DC,1})/\mathcal{R}_{\rm t,1}\simeq 1$. As $M_{\rm BH,1}\rightarrow\widehat{M}_{\rm BH,1}$, the cross sections of FTDEs and DCs become comparable. 
    
    \item $M_{\rm BH,1}\geq\widehat{M}_{\rm BH,1}$, $M_{\rm BH,2}<\widehat{M}_{\rm BH,2}$: the primary BH significantly affects the interactions with the stars, but only the secondary BH is capable of disrupting the stars. So when it comes to FTDEs, the binary will act effectively as a single BH. The FTDEs can be suppressed because the stars can be ejected or directly captured by the primary BH. 
    
    \item $M_{\rm BH,2}\geq\widehat{M}_{\rm BH,2}$: both BHs can only directly capture the stars and, therefore, FTDEs are completely suppressed. 
\end{enumerate} 

On the other hand, $\Sigma_{\rm DC}$ monotonically increases with $M_{\rm BH}$. Therefore, the stars directly plunge into the more massive BH preferentially because of the larger cross section ($\Sigma_{\rm DC, 2}/\Sigma_{\rm DC, 1} \simeq q$). But as explained above, for small BHs, DCs would be significantly suppressed because $\Sigma_{\rm FTDE}/\Sigma_{\rm DC}\gg1$.

The cross section for FTDEs and $\widehat{M}_{\rm BH}$ depends on $M_{\star}$. However, the cross section for DCs is independent of $M_{\star}$.

This simple argument based on the cross section does not take into account the impact of chaotic interactions \citep[e.g.][]{stone19,manwadkar20,manwadkar21,parischewsky21} or secular evolution in triples \citep[e.g.][]{fragione18a,fragione18b,fragione18c} which might be significant. However, as we will show in \S\ref{sec:result}, our results can be reasonably well described using the concept of the cross section.

\begin{figure*}
	\centering
	\includegraphics[width=8.6cm]{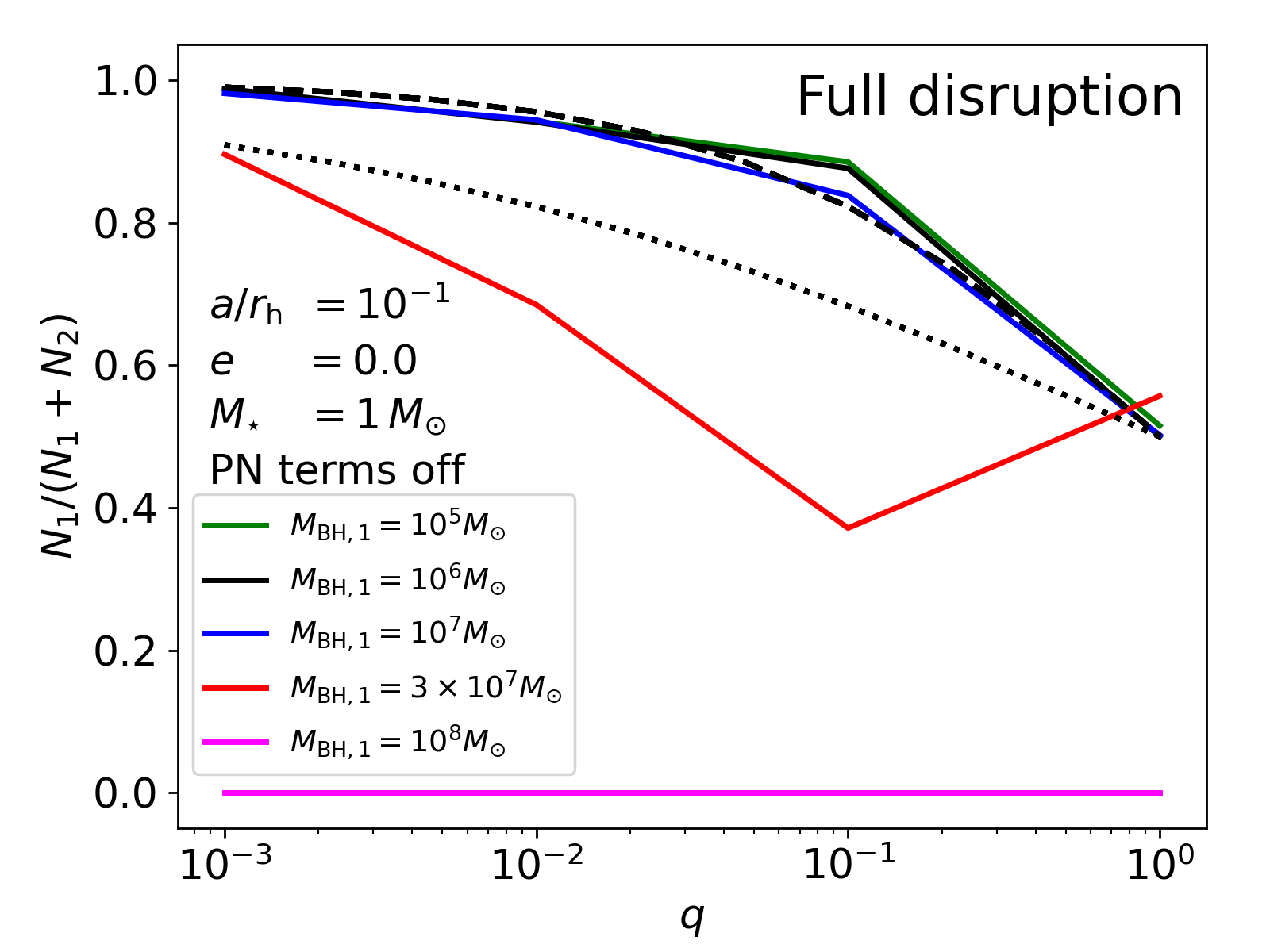}
	\includegraphics[width=8.6cm]{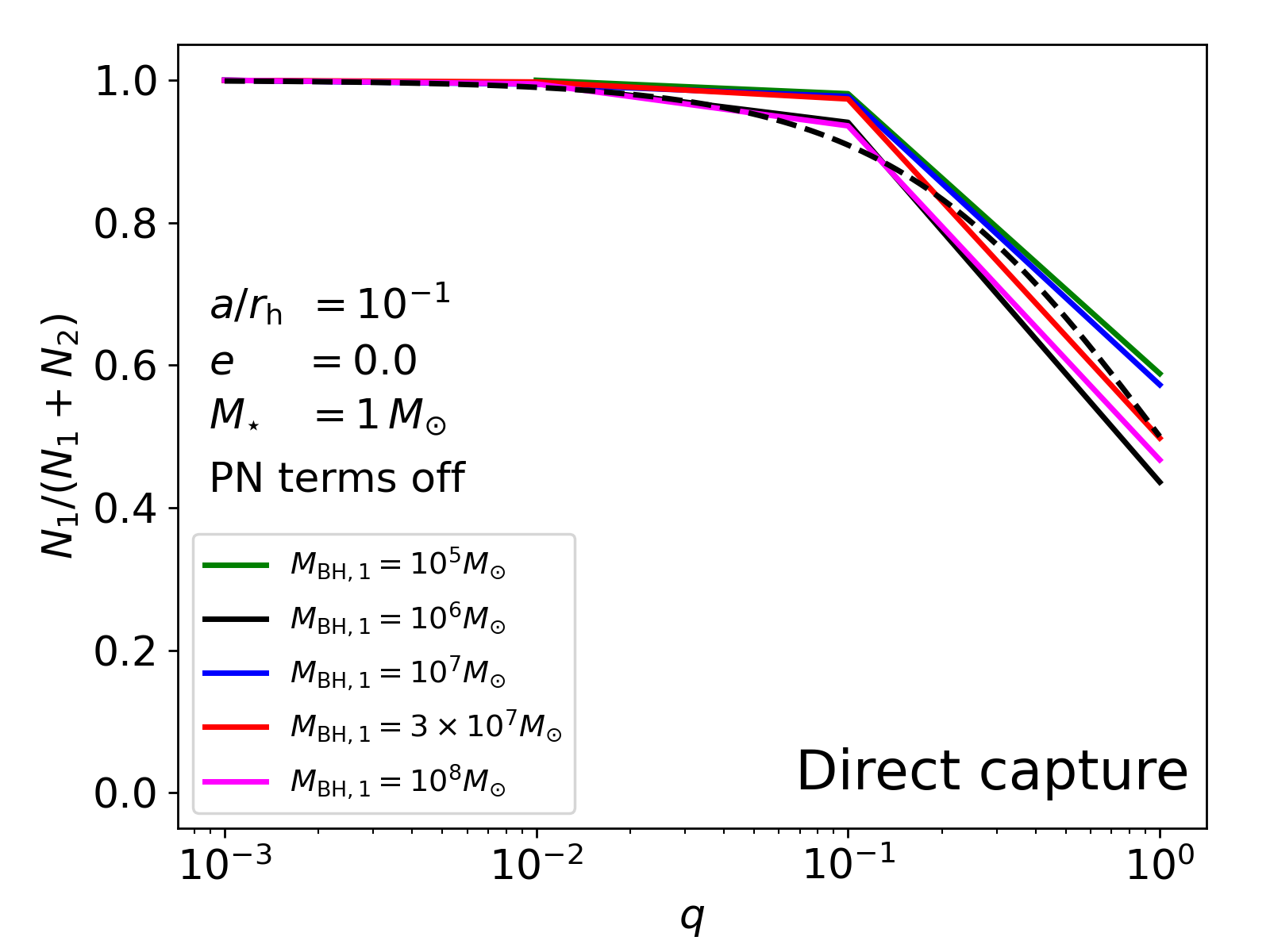}	
\caption{The ratio of the number of events $N_{\rm 1}$ for FTDEs (\textit{left}) and DCs (\textit{right}) for a $1\Msol$ star by the primary BH in circular SMBHBs with $a/r_{\rm h}=10^{-1}$ to the total number of events ($N_{1}+N_{2}$), as a function of $q$, for different values of $\pbhm$. The black dotted lines in both panels indicate the predicted values using analytic expressions with the cross sections: $1/(1+q^{1/3})$ for FTDEs and $1/(1+q)$ for DCs. The analytic expression for DCs gives a reasonably good fit to the event fraction, but that for FTDEs underestimates the event fraction by 10-15\%. So we depict a better fit, $1/(1+q^{1/1.5})$, using a black dashed line.}
	\label{fig:eventfraction_mbh}
\end{figure*}

\section{Results}\label{sec:result}

In this section, we present and analyze the results of our numerical scattering experiments.

\subsection{Event fractions for full disruptions and direct captures}\label{sub:eventfraction}

 We find that the number of DCs $N_{1}$ by the primary black hole relative to the total number of the events $N_{1}+N_{2}$ has a strong dependence on $q$, but only weakly depends on $a$ and $e$, and is nearly independent of $M_{\rm BH,1}$ and $M_{\star}$. The event fraction $N_{1}/(N_{1}+N_{2})$ for FTDEs by SMBHBs with $\pbhm\ll\widehat{M}_{\rm BH,1}$ has a qualitatively similar dependence as that for DCs. However, it has an $\mstar$-dependence associated with the encounter cross section (see Figures~\ref{fig:cross} and \ref{fig:eventfraction_mstar}). When $\pbhm>\widehat{M}_{\rm BH,1}$, the event fraction for FTDEs is zero because the primary BH can not disrupt stars ($N_{1}=0$).

We provide a more detailed description of the dependence on the key parameters in the following subsections.

\subsubsection{Dependence on stellar mass}\label{subsub:event_frac_mstar}
The $\mstar$-dependence of the event fraction for FTDEs and DCs is very well-described by that of the encounter cross section. As an example, we show the event fraction for FTDEs by circular unequal-mass SMBHBs with $a/r_{\rm h}=10^{-2}$ in Figure~\ref{fig:eventfraction_mstar}. For $\pbhm\leq 10^{6}\Msol$, the event fraction is nearly independent of $M_{\star}$. However, as $\pbhm$ approaches $\widehat{M}_{\rm BH,1}\simeq (3-7)\times10^{7}\Msol$, the dependence becomes stronger. Finally when $\pbhm>\widehat{M}_{\rm BH,1}$, the event fraction becomes zero simply because the primary BH can not disrupt stars\footnote{We performed additional simulations for circular binaries $\pbhm=5\times10^{7}\Msol$ which is greater than $\widehat{M}_{\rm BH,1}$ for $\mstar=1\Msol$, but smaller than that for $\mstar=3\Msol$. As expected, $N_{1}/(N_{1}+N_{2})=0$ for $\mstar=1\Msol$. But $N_{1}/(N_{1}+N_{2})$ for $\mstar=3\Msol$ is $\simeq0.6$ at $q=10^{-1}$, $\simeq 0.8$ at $q=10^{-2}$ and $\simeq 0.9$ at $q=10^{-3}$.}. This trend is exactly what is expected from the cross section. To confirm this, we plot the values of $\Sigma_{\mathrm{TDE},1}/(\Sigma_{\mathrm{TDE},1}+\Sigma_{\mathrm{TDE},2})$ + 0.2 for given values of $\pbhm\leq3\times 10^{7}\Msol$ and $\sbhm$ using dashed lines in Figure~\ref{fig:eventfraction_mstar}. Although we shift the lines vertically by $0.2$, the shape of the lines describes the overall trends very accurately.

\begin{figure*}
	\centering
	\includegraphics[width=8.6cm]{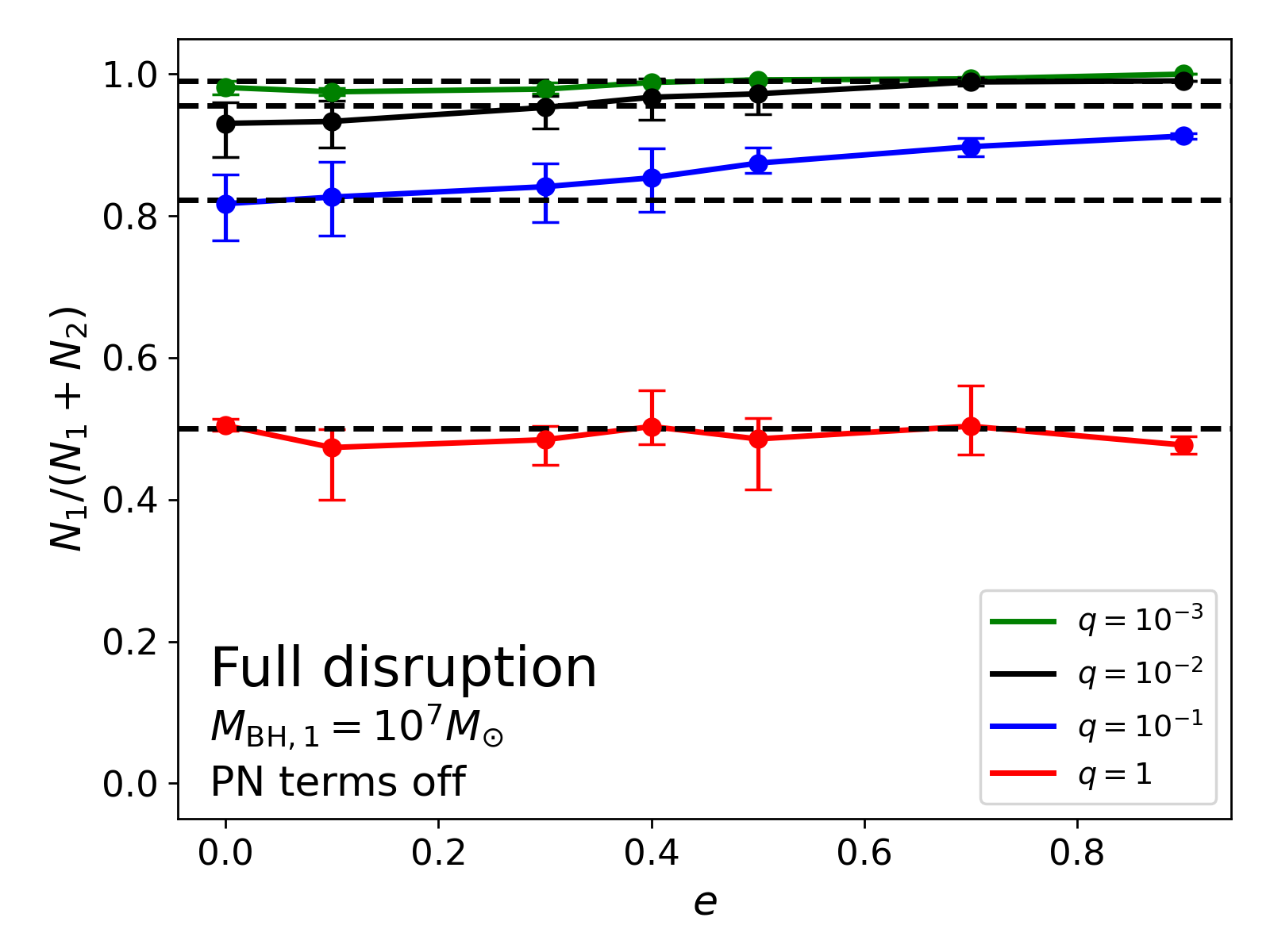}
	\includegraphics[width=8.6cm]{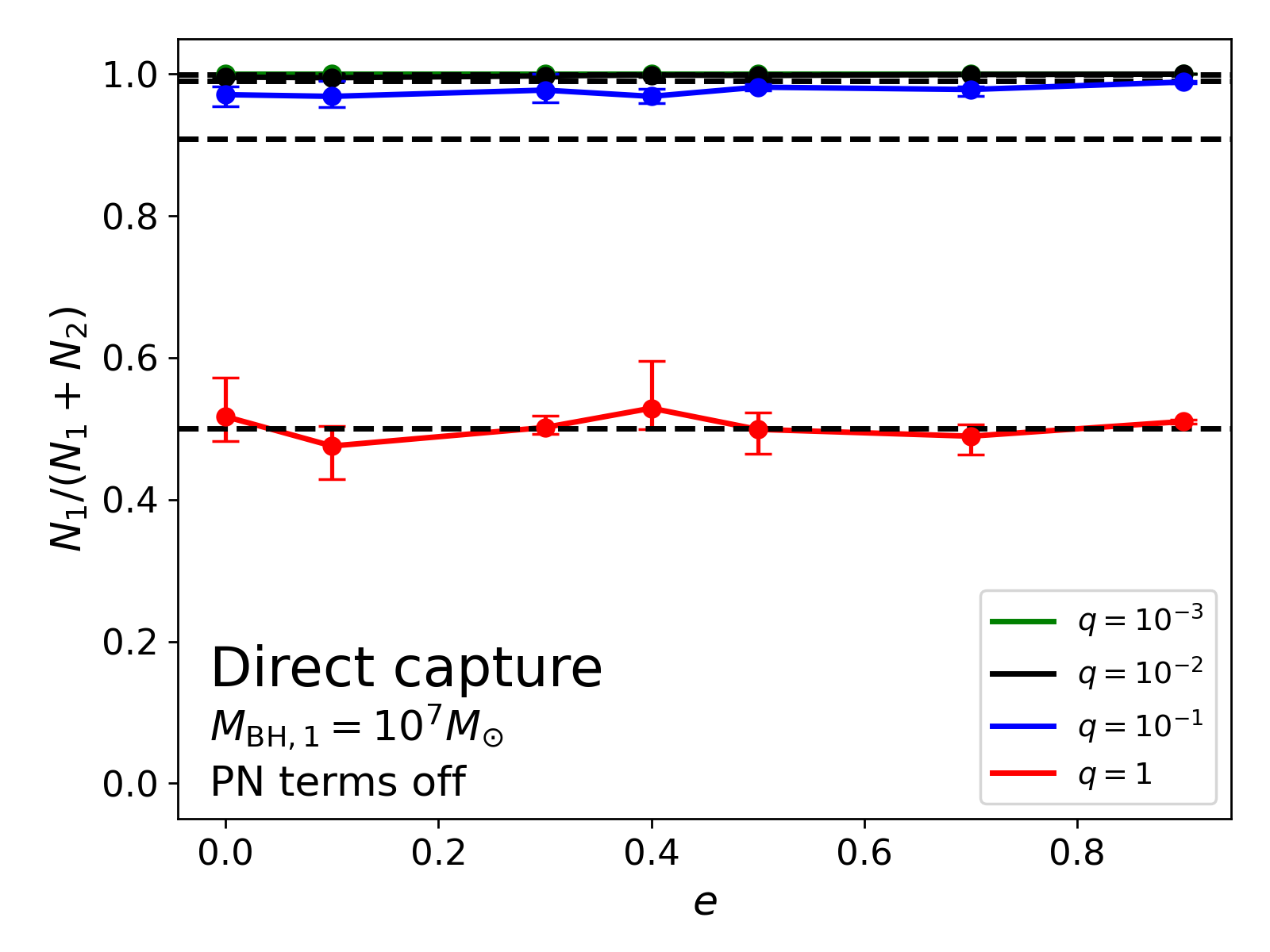}	
\caption{The ratio of the number of events $N_{1}$ for FTDEs (\textit{left}) and DCs (\textit{right}) by the primary BH with $\pbhm=10^{7}\Msol$ to the total number of events $N_{1} + N_{2}$, as a function of $e$, for different values of $q$. The bar at each dot indicates the entire range of variations within the range of the semimajor axes considered for given values of $e$ and $q$. The black dashed lines in both panels indicate the predicted values using the following expressions: $1/(1+q^{1/1.5})$ for FTDEs and $1/(1+q)$ for DCs. }
	\label{fig:eventfraction_ecc}
\end{figure*}

The DC event fraction is nearly independent of $\mstar$, as expected from their cross section.

Because we could not find any significant $\mstar-$dependence deviating from that of the encounter cross section and event probability, without loss of generality, we will present our results for $\mstar=1\Msol$ unless stated otherwise in the rest of the paper.

\subsubsection{Dependence on black hole mass }
The event fraction for both FTDEs and DCs increases as $q$ decreases. This trend can be understood because of the relatively large cross section of the primary BH for low $q$.

We show in Figure~\ref{fig:eventfraction_mbh} the event fraction $N_{1}/(N_{1}+N_{2})$ for FTDEs (\textit{left} panel) and DCs (\textit{right} panel) by circular SMBHBs with $a/r_{\rm h}=10^{-1}$ as a function of $q$ for different values of $\pbhm$. 
The two BHs in equal-mass binaries are equally responsible for the both star-removing events, i.e., $N_{1}/(N_{1}+N_{2})=0.5$. However, as $q$ decreases, the primary BH dominantly destroys the stars. 

For $\pbhm\leq 10^{7}\Msol$, around $80-90\%$ of FTDEs are produced by the primary BH in binaries with $q= 10^{-1}$ and $\gtrsim 90\%$ in binaries with $q<10^{-1}$. This is consistent with the same quantity for circular binaries with $\pbhm=10^{6}\Msol$ from \citet{Darbha+2018} and \citet{fragione18b}. The cross section ratio for FTDEs, i.e., $\Sigma_{\rm FTDE, 1}/(\Sigma_{\rm FTDE, 1}+\Sigma_{\rm FTDE, 2})\simeq 1/(1+q^{1/3})$ for $\pbhm\ll\widehat{M}_{\rm BH,1}$, describes the trend on a qualitative level (but underestimates the event fraction by $\simeq 10-15\%$), as depicted using a black dotted line in the \textit{left} panel. In fact, it is better described by  $N_{1}/(N_{1}+N_{2})=1/(1+q^{1/1.5})$ (the black dashed line). 

The event fraction for FTDEs falls as  $\pbhm>10^{7}\Msol$ and finally becomes zero at $\pbhm=10^{8}\Msol$. Interestingly, the event fraction for $\pbhm=3\times10^{7}\Msol$ is not monotonically decreasing as $q$ increases. We attribute this to the competition between the deeper potential of the primary BH affecting the stellar orbit more and the larger cross section of the secondary BH, i.e., $\Sigma_{\rm FTDE,2}\geq\Sigma_{\rm FTDE,1}$.

The event fraction for DCs converges to unity at larger $q$ than FTDEs. More than 90\% of the events are created by the primary BH in binaries with $q=10^{-1}$. The analytic expression for the cross section ratio $\Sigma_{\rm DC, 1}/(\Sigma_{\rm DC, 1}+\Sigma_{\rm DC, 2})= 1/(1+q)$ (the black  dashed line in the \textit{right} panel of Figure~\ref{fig:eventfraction_mbh}) provides a good description of the event fraction for DCs.

\subsubsection{Dependence on semimajor axis and eccentricity }
The dependence of $N_{1}/(N_{1}+N_{2})$ on $a/r_{\rm h}$ and $e$ are both relatively weak. For $10^{-3}<q\leq1$, $N_{1}/(N_{1}+N_{2})$ rises with $a/r_{\rm h}$ and $e$, which only results in variations by $\lesssim 10-20\%$ for FTDEs and $\lesssim 5\%$ for DCs across the range of  $10^{-3}\leq a/r_{\rm h}\leq 1$ and $0\leq e\leq 0.9$ for any given $q$. However, for $q= 10^{-3}$ or $1$, the event fraction does not depend on $a/r_{\rm h}$ and $e$ any more because it asymptotes to a single value, e.g., 1 for $q= 10^{-3}$ or 0.5 for $q=1$. As an example, we show in Figure~\ref{fig:eventfraction_ecc} the event fraction as a function of $e$ for FTDEs (\textit{left}) and DCs (\textit{right}) by the primary mass $\pbhm=10^{7}\Msol$. The bars at each dot indicate the entire variation across the range of $a/r_{\rm h}$ considered. As just mentioned, the event fraction for FTDEs monotonically increases with $e$ for $q=10^{-1}$ and $10^{-2}$ while it remains constant at $\simeq 0.5$ for $q=1$ and $\simeq1$ for $q=10^{-3}$. The event fraction for DCs shows similar trends, but more steeply changes with $q$ and more weakly depends on $a$ and $e$ than do FTDEs. Note that the eccentricity not only increases the event fraction but also the event probability, which will be presented in \S~\ref{subsub:ecc}.

\subsection{The enhancement and suppression of disruptions by supermassive black hole binaries}
In this section, we investigate the probability of FTDEs by SMBHBs relative to that by single SMBHs of the same mass ($\pbhm+\sbhm$). We define the relative probability $\Upsilon$ for FTDEs as,
\begin{align}\label{eq:rel_proba}
    \UpTDE &=\left(\frac{N_{\rm FTDE}}{N_{\rm tot}}\right)\left(\frac{\mathcal{R}_{\rm t}-\mathcal{R}_{\rm DC}}{2a}\right)^{-1},
\end{align}
where $N_{\rm FTDE}$ is the number of FTDEs and $N_{\rm tot}$ is the total number of encounters. Here, $\mathcal{R}_{\rm t}$ and $\mathcal{R}_{\rm DC}$ are the maximum value of the quantity between the primary and secondary BHs.  Note that our definition of $\UpTDE$ is different from that of \citet{Darbha+2018} -- they define it using the relative cross section for both FTDEs and DCs which is expressed in terms of the canonical tidal radius for the primary BH, i.e., $r_{\rm t,1}/2a$. 

\begin{figure}
	\centering
	\includegraphics[width=7.6cm]{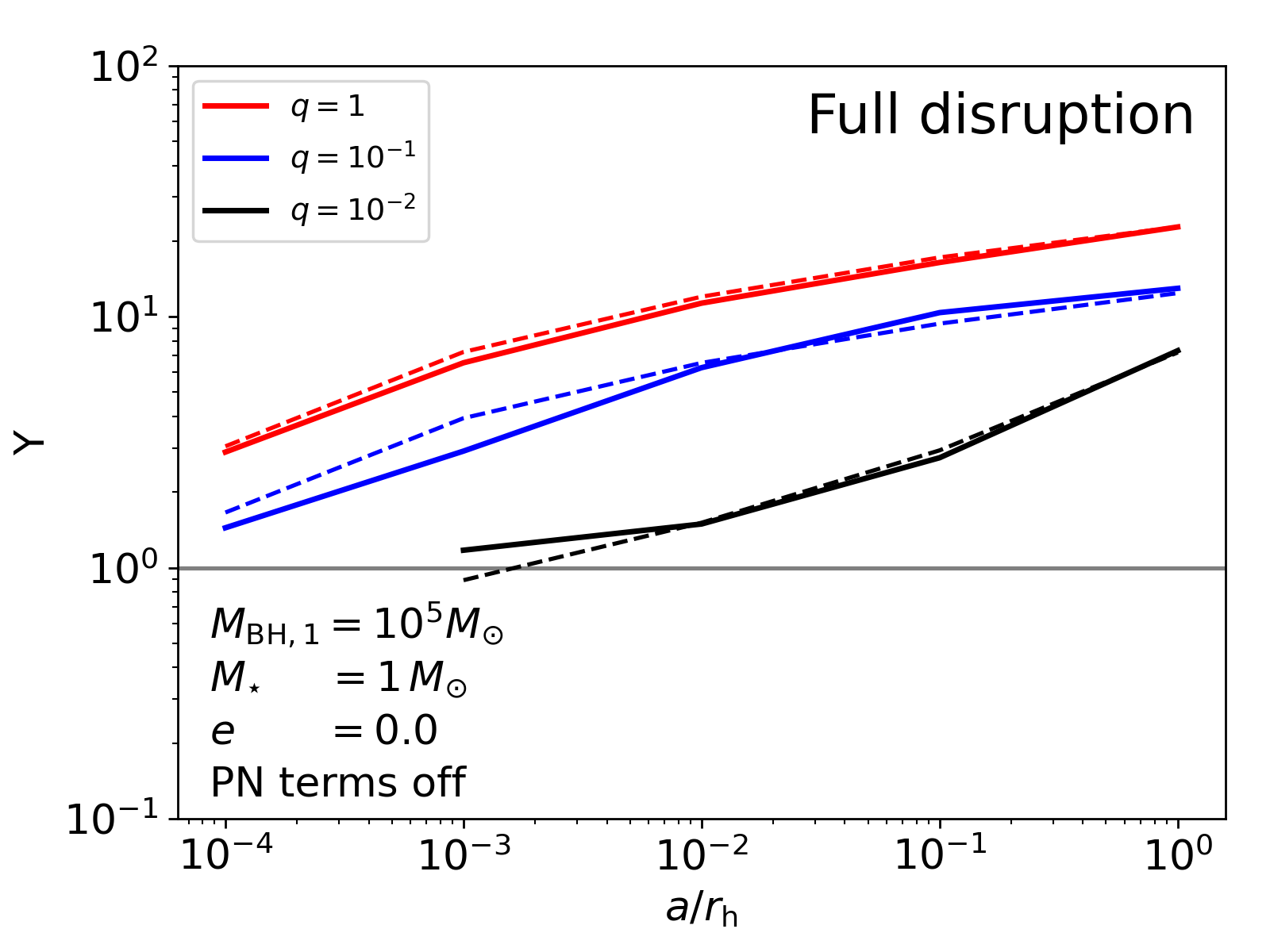}\vspace{-0.1in}
	\includegraphics[width=7.6cm]{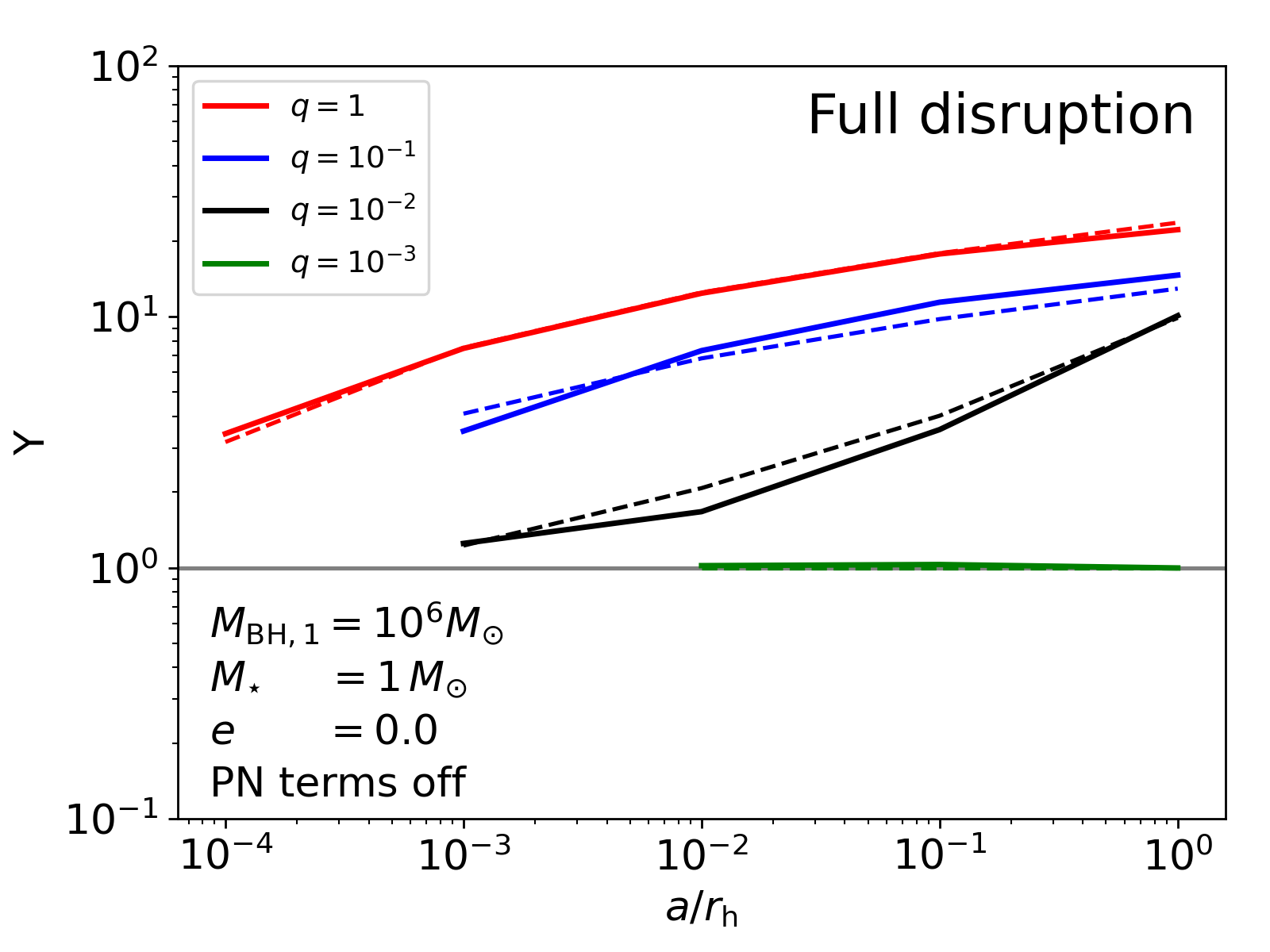}\vspace{-0.1in}
	\includegraphics[width=7.6cm]{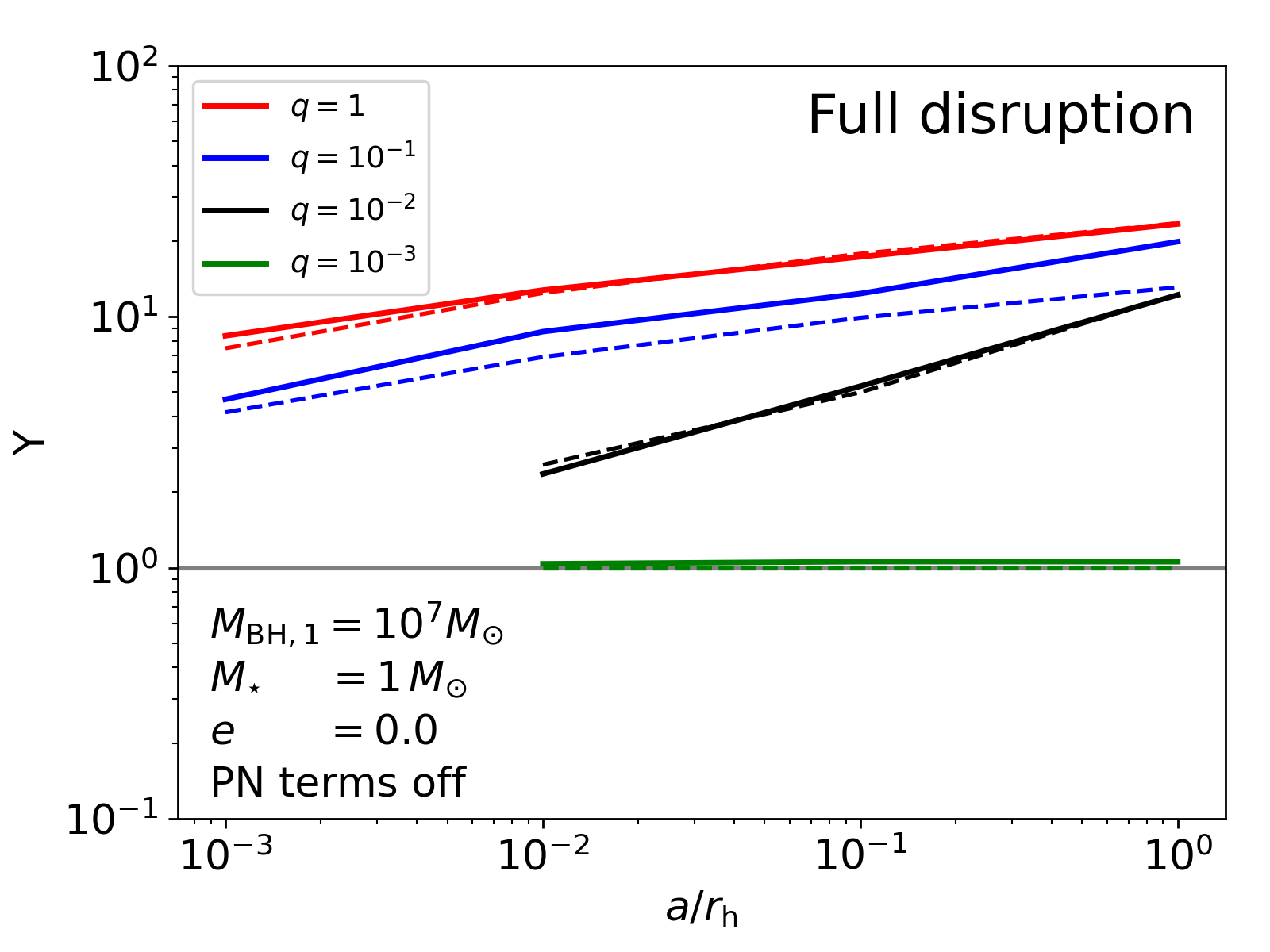}\vspace{-0.1in}
	\includegraphics[width=7.6cm]{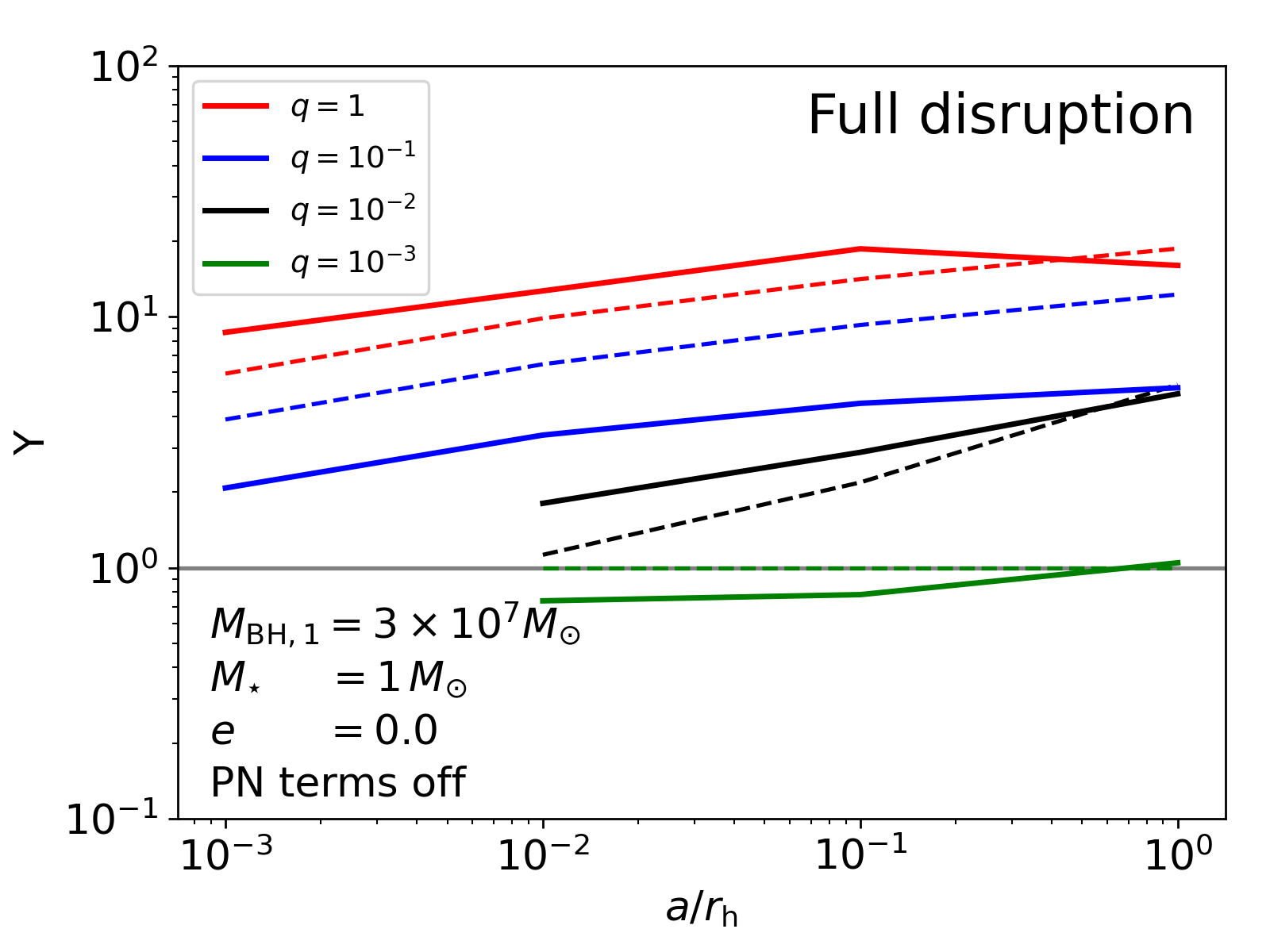}\vspace{-0.1in}	
\caption{The relative probability $\Upsilon_{\rm TDE}$ (Equation~\ref{eq:rel_proba}) for full disruptions by circular binaries with $\pbhm=10^{5}$, $10^{6}$, $10^{7}$ and $3\times10^{7}\Msol$ as a function of $a/r_{\rm h}$. The dashed lines show the predicted values from our fitting formula (Equation~\ref{eq:upTDE_fit}). }
	\label{fig:tde1}
\end{figure}

\begin{figure}
	\centering
	\includegraphics[width=8.6cm]{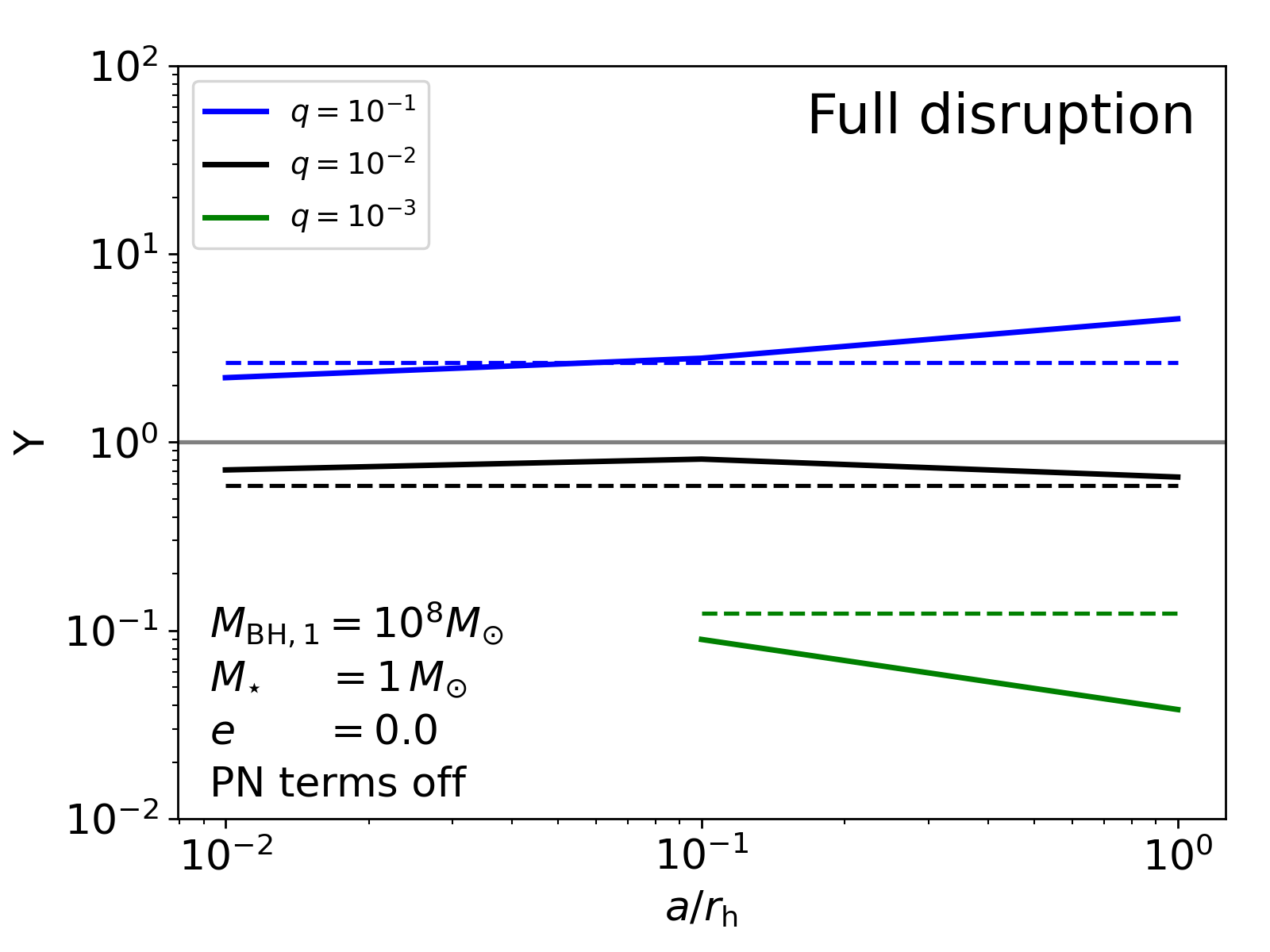}
\caption{The same as Figure~\ref{fig:tde1}, but for a binary with $\pbhm=10^{8}\Msol$. The dashed lines show the predicted values estimated using our fitting formula (Equation~\ref{eq:upTDE_fit}).}
	\label{fig:tde2}
\end{figure}

\begin{figure*}
	\centering
	\includegraphics[width=8.6cm]{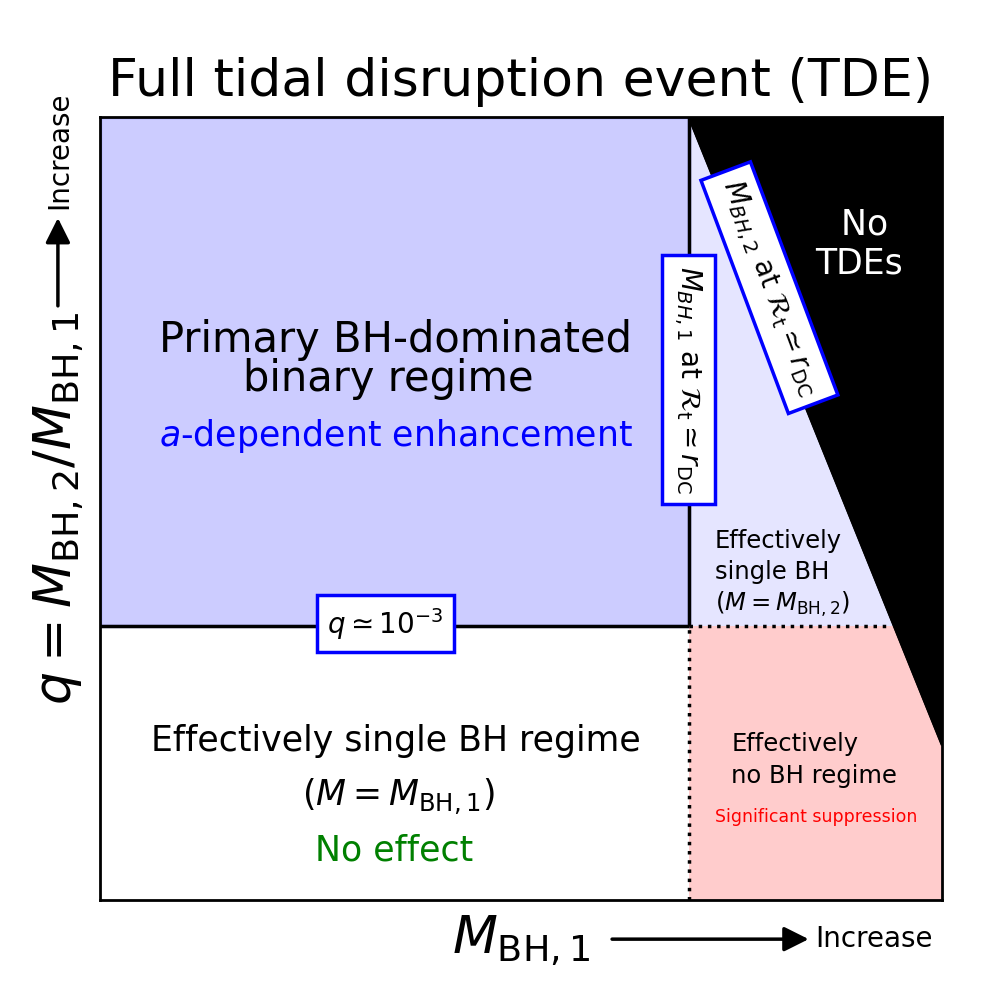}
	\includegraphics[width=8.6cm]{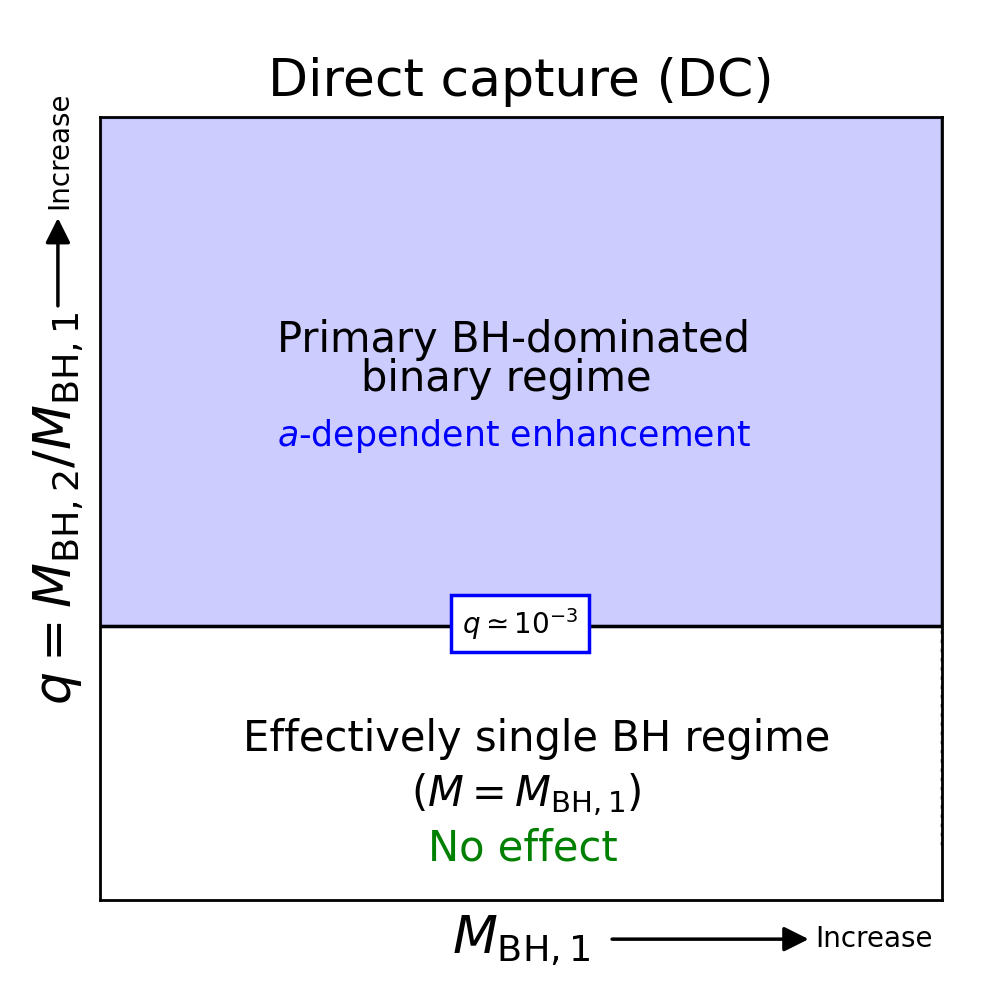}\caption{Schematic diagrams for the dependence of $\Upsilon$ on $M_{\rm BH}$ for full disruptions (\textit{left}) and direct captures (\textit{left}). Different colors indicate roughly how much SMBHBs 
	enhance the event rate relative to single SMBHs: significant enhancement (darker blue color), either slight or little to no enhancement (lighter blue), no binary effect (white), significant suppression (red) and no TDEs by either one of the BHs (black). 
	\textit{Left}: When $q>10^{-3}$ and $\pbhm<\widehat{M}_{\rm BH,1}$, the events are dominated by the primary BH and the existence of the secondary BH enhances the event probability (i.e., the primary BH-dominated binary regime). In this regime, $\Upsilon$ increases with $a$. However, for those binaries with $\pbhm>\widehat{M}_{\rm BH,1}$, the event probability is primarily determined by the secondary BH (i.e., effectively the single BH regime with $M_{\rm BH}=M_{\rm BH,2}$). Even when $\pbhm<\widehat{M}_{\rm BH,1}$, the binary with $q\lesssim 10^{-3}$ acts as a single BH with $M_{\rm BH}=\pbhm$ because the perturbation of the secondary BH is so weak. Lastly, FTDEs are significantly suppressed around SHBMHs with $q\lesssim 10^{-3}$ and $\pbhm>\widehat{M}_{\rm BH,1}$. (\textit{Right}): SMBHBs only enhance ($q>10^{-3}$) or have no significant impact ($q\lesssim10^{-3}$) on the DC rates. }
	\label{fig:parameter}
\end{figure*}

\begin{table*}
    \centering
	\begin{tabular}{c c c c c}
		\hline
		\multicolumn{5}{c}{Full tidal disruption event}\\
		\hline
		 Regime &  Primary BH-dominated binary  & Effectively single ($M=\pbhm$) & Effectively single ($M=\sbhm$)  & Effectively no BH \\
		$\pbhm$  & $\pbhm<\widehat{M}_{\rm BH,1}$ & $\pbhm<\widehat{M}_{\rm BH,1}$ & $\pbhm>\widehat{M}_{\rm BH,1}$ & $\pbhm>\widehat{M}_{\rm BH,1}$   \\ 
		 $\sbhm$   & $\sbhm<\widehat{M}_{\rm BH,2}$ & $\sbhm<\widehat{M}_{\rm BH,2}$ & $\sbhm<\widehat{M}_{\rm BH,2}$ & $\sbhm<\widehat{M}_{\rm BH,2}$   \\ 
		$q$   & $>10^{-3}$ & $\lesssim10^{-3}$ & $>10^{-3}$ &  $\lesssim10^{-3}$\\
		\hline
		$M_{\rm BH}-$dep. & $\UpTDE\propto q^{\alpha}$ with $\alpha\simeq 0.25-0.3$ & No dependence & No dependence & No dependence  \\
		$M_{\star}-$dep. & \multicolumn{4}{c}{$\propto \Sigma_{\rm TDE}\propto \mathcal{R}_{\rm t}-\mathcal{R}_{\rm DC}$}\\
		$a-$dep. & increases with $a/r_{\rm h}$ & No dependence & No dependence& No dependence \\
		$e-$dep. &increases with $e$  & No dependence & No dependence & No dependence \\
    \hline
 	\end{tabular}
		\caption{Four regimes of full tidal disruption events depending on $\pbhm$ and $\sbhm$, primary BH-dominated binary, effectively single black hole with $M=\pbhm$, effectively single BH with $M=\sbhm$ and effectively no BH, and the dependence of key parameters, $M_{\rm BH}$, $M_{\star}$, $q$ and $e$. For direct captures, there are only two regimes, primary BH-dominated binary and effectively single BH with $M=\pbhm$, whose dependence on the key parameters is qualitatively the same with the same regime for FTDEs. See also Figure~\ref{fig:parameter}. } \label{tab:TDEregimes}
\end{table*}

\begin{figure*}
	\centering
	\includegraphics[width=8.6cm]{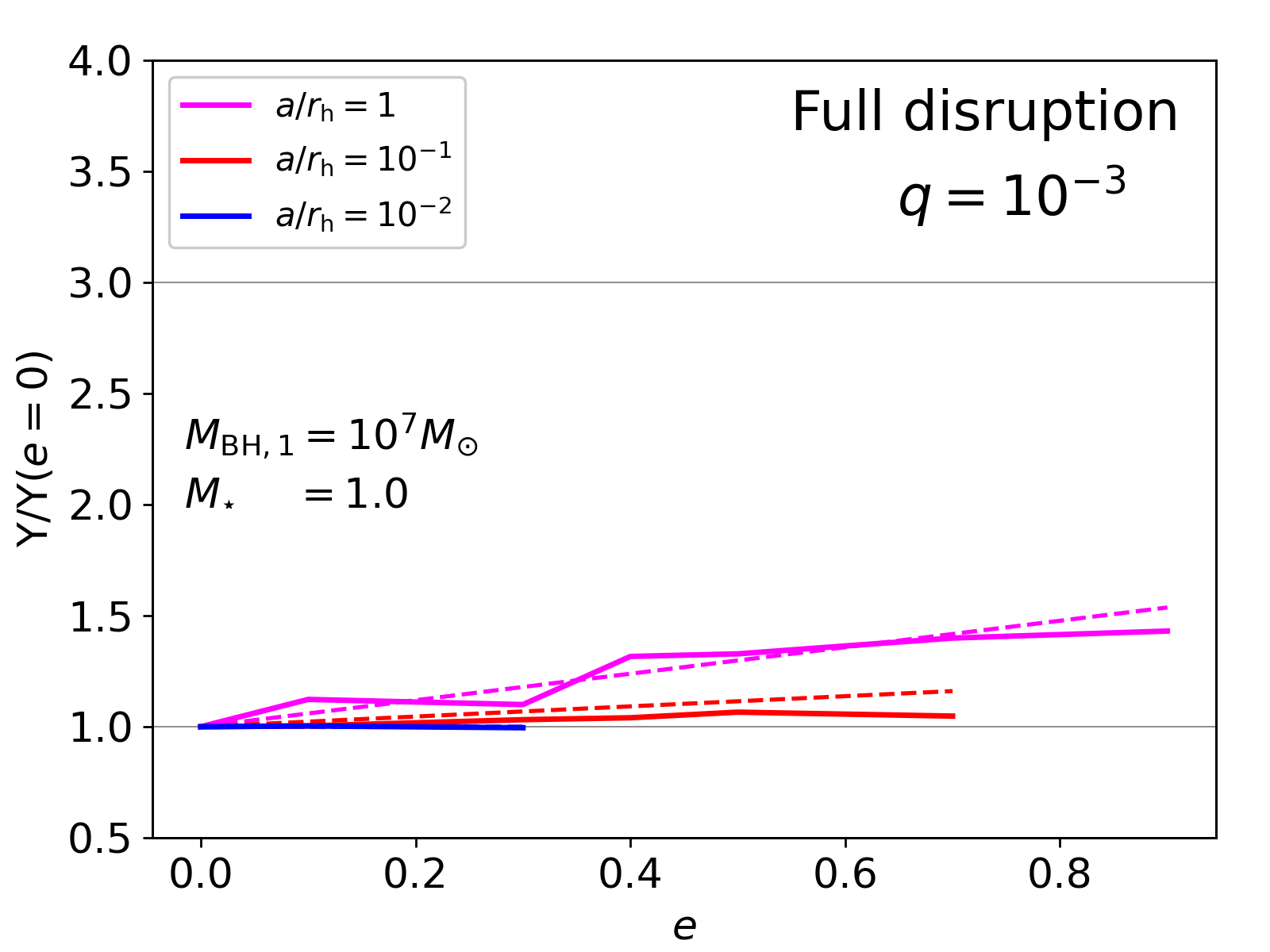}
	\includegraphics[width=8.6cm]{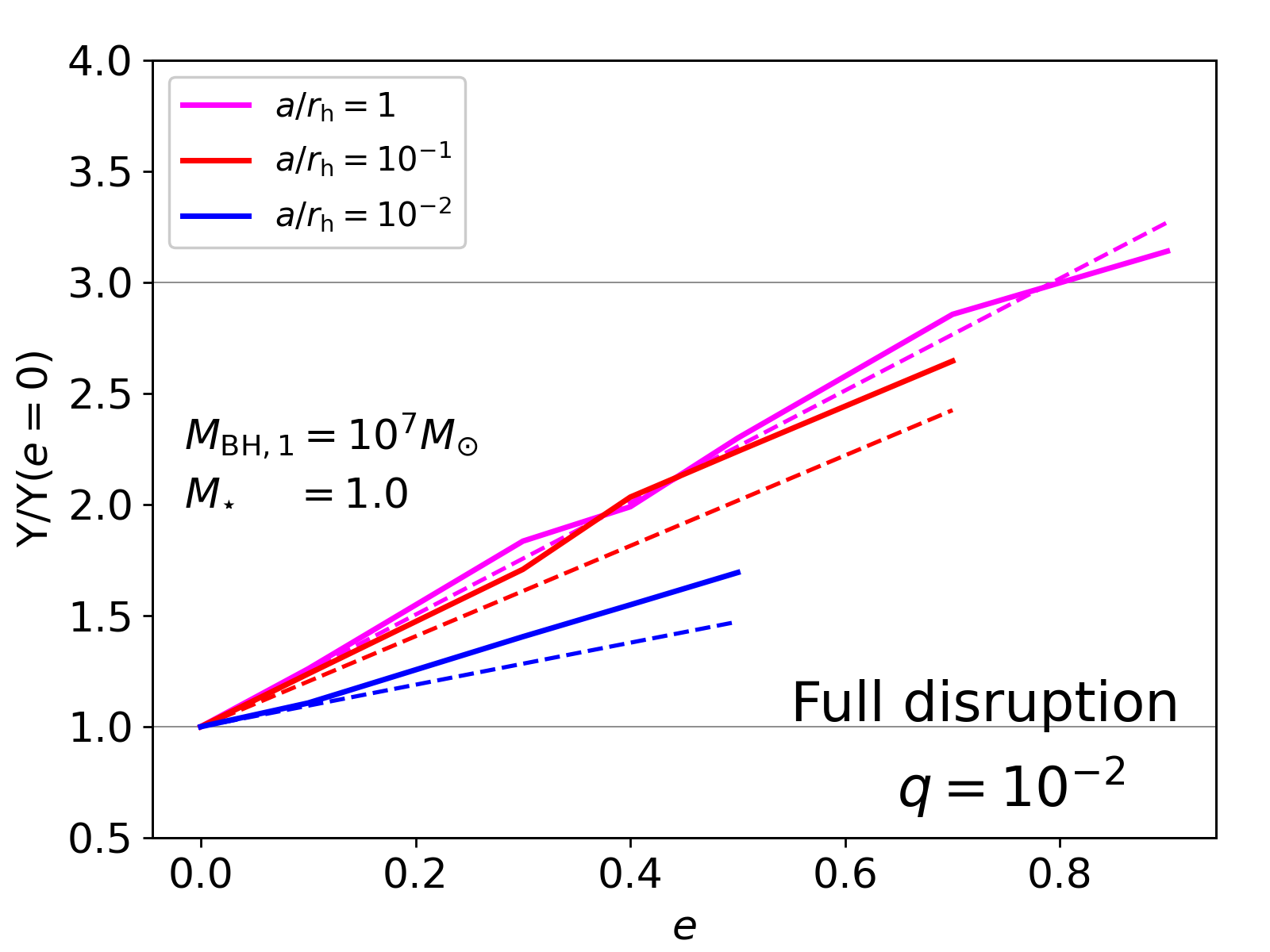}\\
	\includegraphics[width=8.6cm]{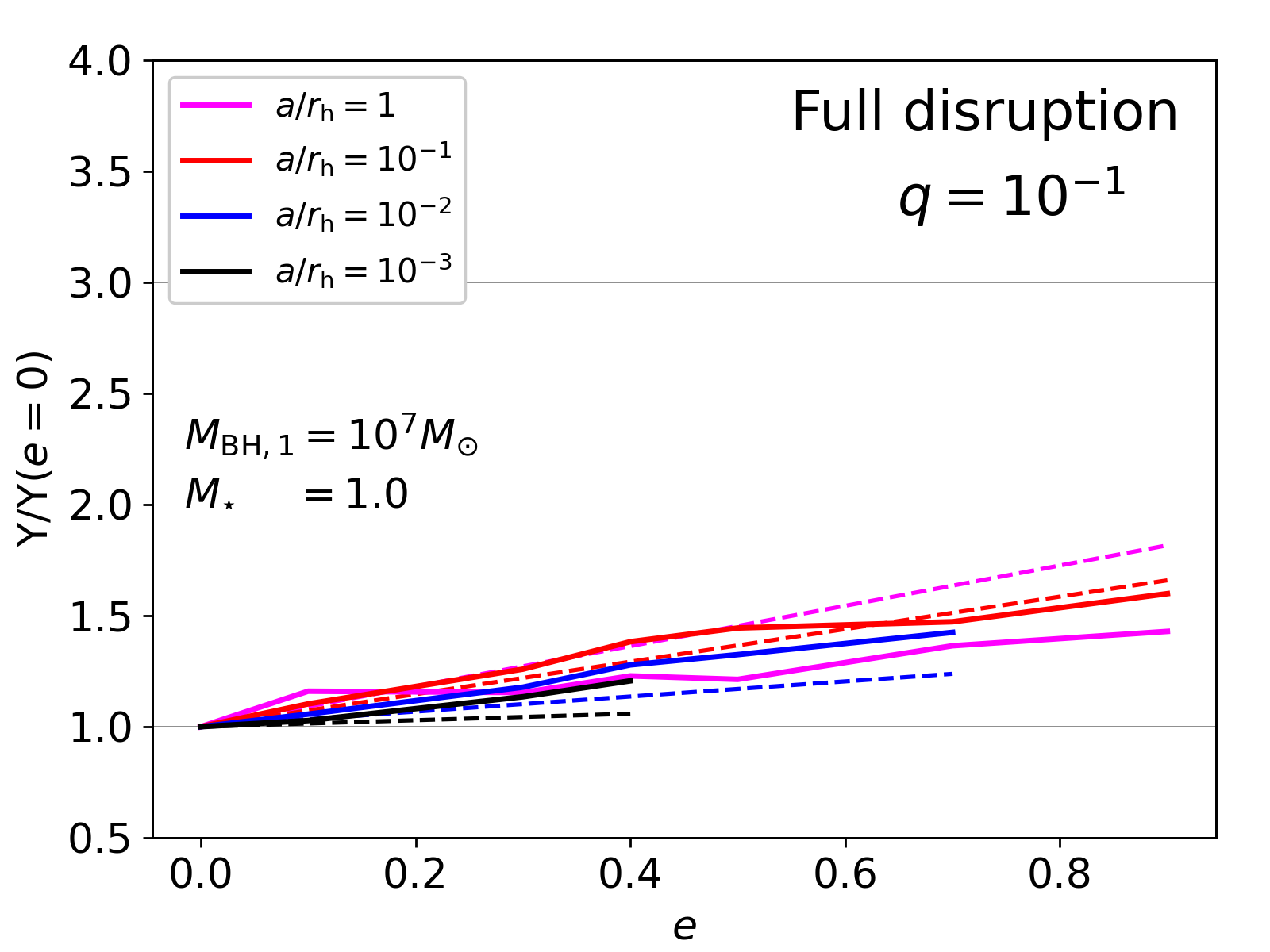}	
	\includegraphics[width=8.6cm]{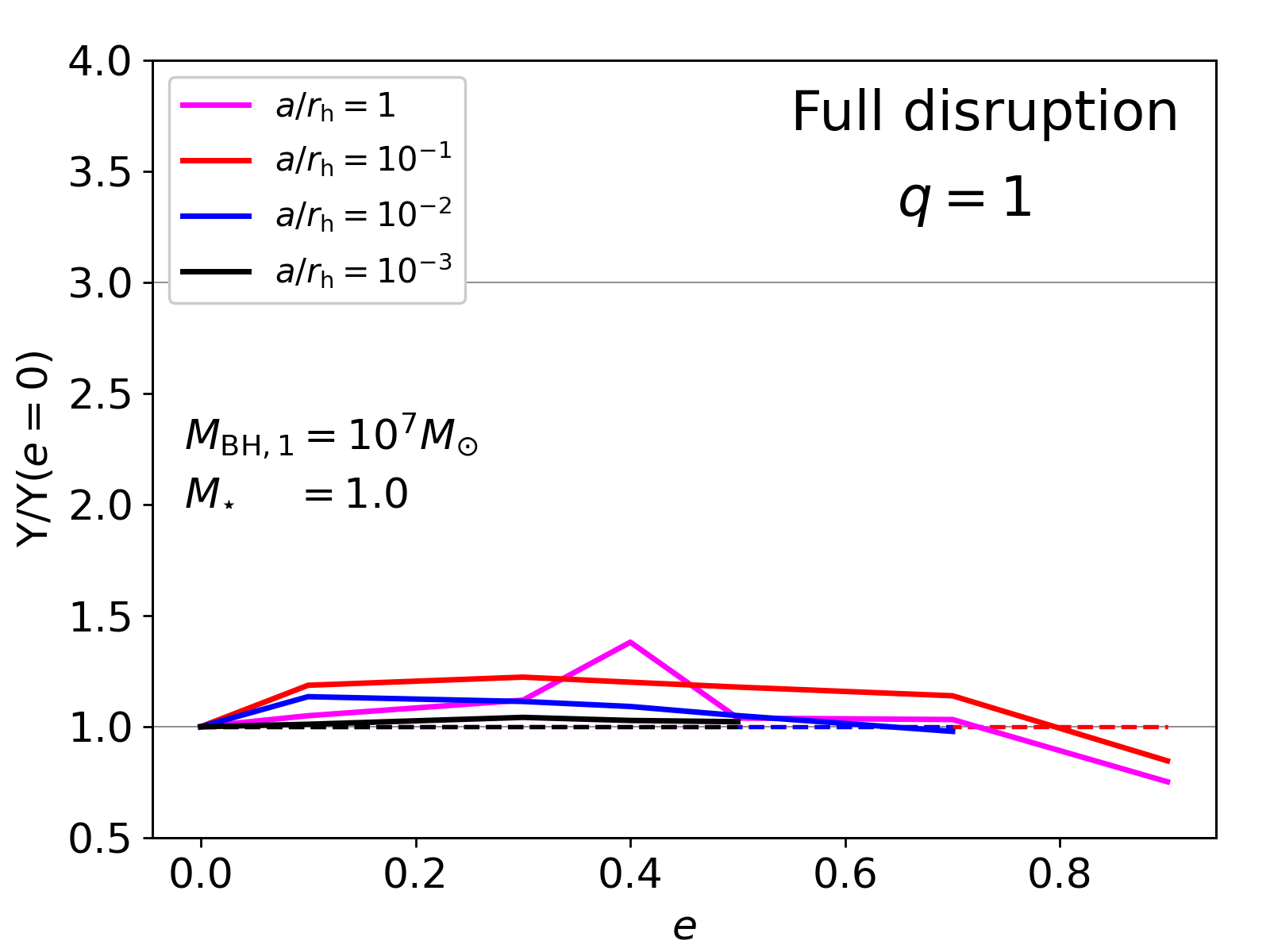}\caption{The relative probability $\Upsilon$, normalized by that for circular binaries for full disruption events as a function of $e$ for $q=10^{-3}$ (\textit{top-left}), $10^{-2}$ (\textit{top-right}), $10^{-1}$ (\textit{bottom-left}) and $1$ (\textit{bottom-right}) when no relativistic effects are included. The dashed lines show the predicted values using our fitting formula for the $e-$boosting effect (Equation~\ref{eq:ecc_boost}).  }
	\label{fig:ecc_dep}
\end{figure*}

\subsubsection{Dependence on black hole mass}

The BH mass dependence of $\UpTDE$ is qualitatively different depending on $\pbhm$ and $q$, which we describe as follows.
\begin{enumerate}
    \item \textit{Primary BH-dominated binary regime} ($q>10^{-3}$ and $\pbhm<\widehat{M}_{\rm BH,1}$)
    
The primary black hole  dominantly disrupts the stars because of its larger cross section (see \S\ref{sub:eventfraction}). Nonetheless, the secondary BH plays an important role in enhancing the probability by providing a large cross section ($\sim2a$) to deflect the incoming star, which otherwise would have passed by if there were only a single SMBH instead of a SMBHB. The subsequent interactions are violent, with significant perturbations. Figure~\ref{fig:tde1} shows $\Upsilon_{\rm TDE}$ for circular SMBHBs with (from \textit{top} to \textit{bottom}) $M_{\rm BH,1}=10^{5}$, $10^{6}$, $10^{7}$ and $3\times10^{7}\Msol$, as a function of $a/r_{\rm h}$. The probability of FTDEs by SMBHBs can be larger by a factor of $\simeq$20 relative to that for single SMBHs. Furthermore, $\UpTDE$ increases with $q$, roughly following $\propto q^{\alpha}$ where $\alpha\simeq 0.25-0.3$, for given values of $M_{\rm BH,1}$ and $a/r_{\rm h}$. Clearly, the enhancement is affected by how compact the binary is (see \S\ref{subsub:semi}).

The $q$-dependence of $\UpTDE$ is qualitatively consistent with \citet{Darbha+2018} in which they consider FTDEs of $1\Msol$ stars by circular binaries with  $M_{\rm BH,1}=10^{6}\Msol$ and $0.01\leq q\leq 1$. However, their $\UpTDE$ is generally smaller by a factor of $1.5-2$ relative to ours for given values of $q$ and $a$.  This is likely because we use different values for the tidal disruption radii and relative cross-sections than for single SMBHs. \\

\item \textit{Single BH regime with $M_{\rm BH}=M_{\rm BH,1}$} ($q\leq 10^{-3}$,  $\pbhm<\widehat{M}_{\rm BH,1}$)

In this regime, the presence of the secondary black hole has no significant impact on the event probability, i.e., $\UpTDE\simeq 1.0$ for almost all ranges of the parameters considered (horizontal lines for $q=10^{-3}$ in Figure~\ref{fig:tde1}). \\

\item \textit{Single BH regime with $M_{\rm BH}=M_{\rm BH,2}$} ($q>10^{-3}$,  $\pbhm>\widehat{M}_{\rm BH,1}$, $\sbhm<\widehat{M}_{\rm BH,2}$)

The binaries act like a single BH of mass $M_{\rm BH}=\sbhm$ when the primary BH can not disrupt stars ($\pbhm>\widehat{M}_{\rm BH,1}$), but the secondary BH still can ($\sbhm<\widehat{M}_{\rm BH,2}$). This is an interesting difference for FTDEs relative to what has been found for single SMBHs. In this case, FTDEs can be at most slightly enhanced. Figure~\ref{fig:tde2} shows $\UpTDE$ for circular SMBHBs with $M_{\rm BH,1}=10^{8}\Msol$ for which $M_{\rm BH,1}>\widehat{M}_{\rm BH,1}$. $\UpTDE$ decreases with $q$: $\UpTDE\simeq 2-3$ for $q=10^{-1}$ and $\UpTDE\simeq 1$ for $q=10^{-2}$. However, one difference is that $\UpTDE$ becomes nearly independent of $a$ (see \S\ref{subsub:semi}). Note that the events are completely suppressed for $q=1$ because $\sbhm>\widehat{M}_{\rm BH,2}$. \\

\item \textit{No BH regime} ($q\leq 10^{-3}$,  $\pbhm>\widehat{M}_{\rm BH,1}$, $\sbhm<\widehat{M}_{\rm BH,2}$)

The TDEs are significantly suppressed because the primary BH dominantly eats the stars before being disrupted by the secondary BH. As shown in Figure~\ref{fig:tde2} (green line), $\UpTDE<10^{-1}$ for $\pbhm=10^{8}\Msol$ and $q=10^{-3}$. 

\end{enumerate}

The black hole mass dependence for FTDEs is summarized in Table~\ref{tab:TDEregimes} and  the \textit{left} panel of Figure~\ref{fig:parameter}.

\begin{figure*}
	\centering
	\includegraphics[width=8.6cm]{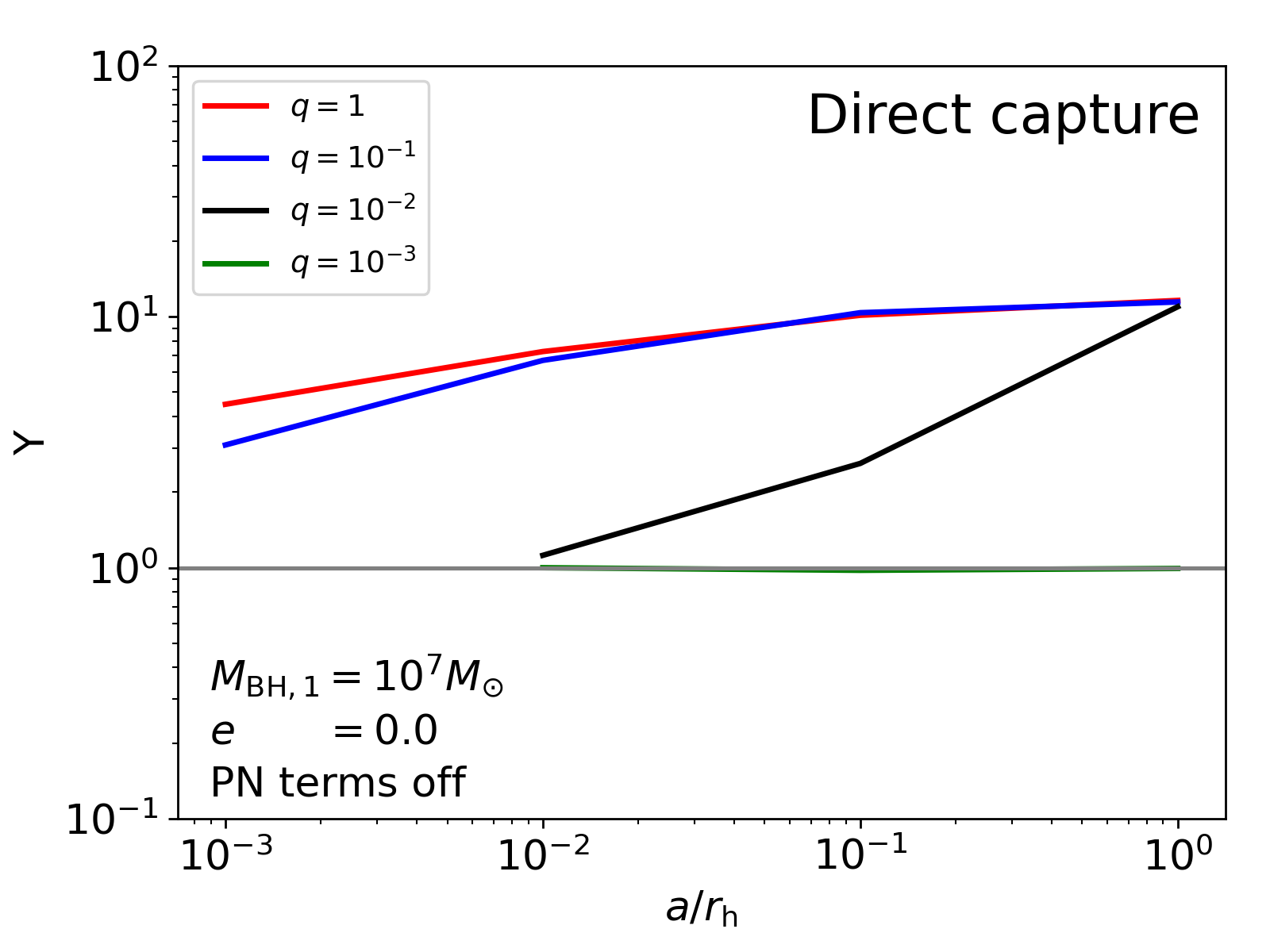}
	\includegraphics[width=8.6cm]{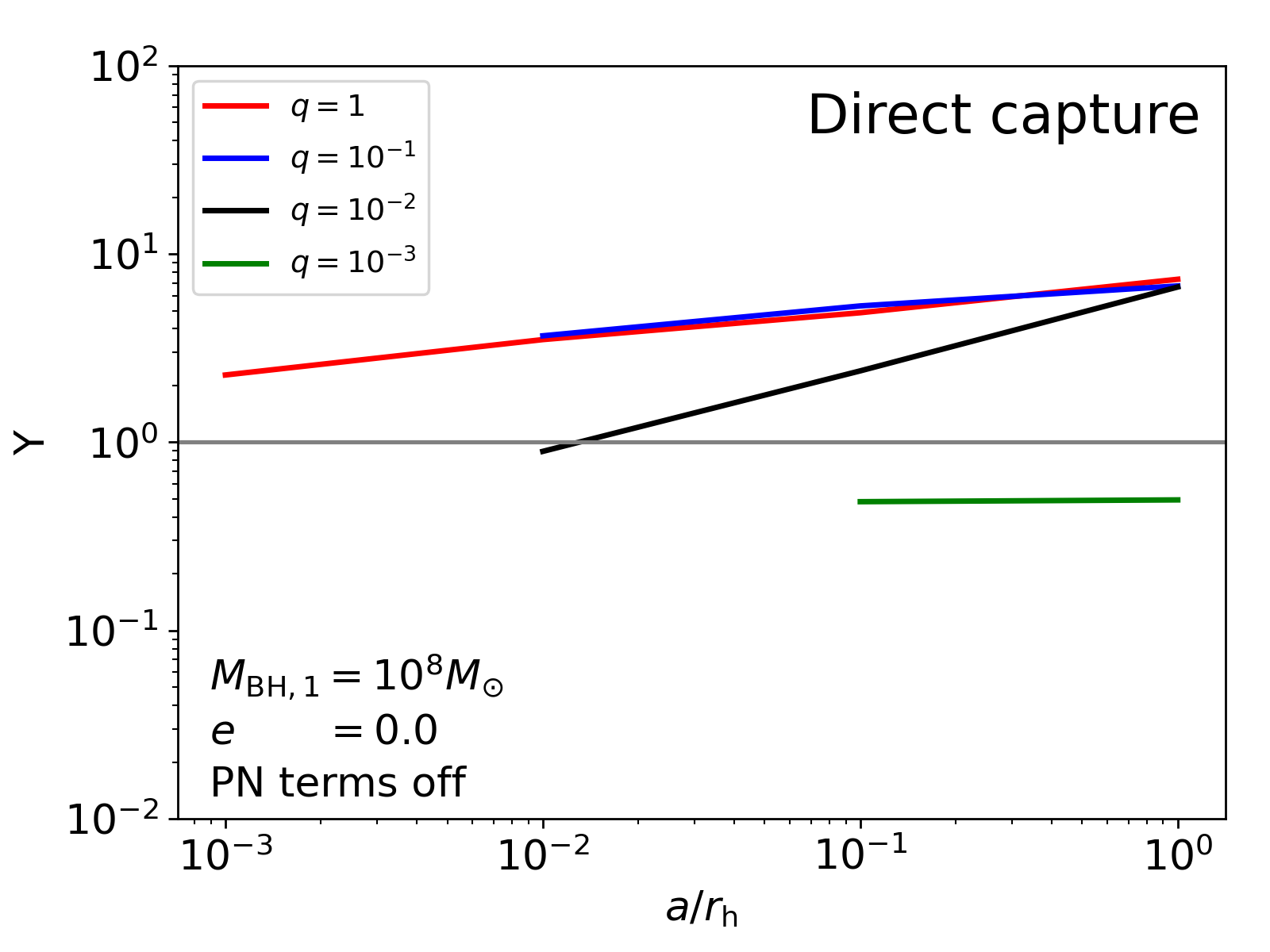}\caption{Same as Figure~\ref{fig:tde1}, but for direct captures by binaries with  $\pbhm=10^{7}\Msol$ (\textit{left}) and  $10^{8}\Msol$ (\textit{right}).}
	\label{fig:dc1}
\end{figure*}

\subsubsection{Dependence on stellar mass}\label{subsub:bhmass}

The dependence of $\UpTDE$ on $\mstar$ is inherited from the encounter cross section, $\propto \mathcal{R}_{\rm t}-\mathcal{R}_{\rm DC}$. That is, it is very weakly dependent on $\mstar$ for $M_{\rm BH}\ll\widehat{M}_{\rm BH}$ but the $M_{\star}$-dependence becomes only noticeable when $M_{\rm BH} \simeq \widehat{M}_{\rm BH}$.

\subsubsection{Dependence on semimajor axis} \label{subsub:semi}
$\Upsilon_{\rm TDE}$ for SMBHBs in the primary BH-dominated binary regime only depends on $a$. In this regime, $\UpTDE$ grows with $a/r_{\rm h}$, which is illustrated in Figure~\ref{fig:tde1}.
It appears that $(\partial\UpTDE/\partial a)(a/\UpTDE)$ also depends on $q$: it increases as $q$ decreases. Overall, $\UpTDE$ for circular binaries varies by factors of 3-20 within the range $10^{-4}\leq a/r_{\rm h}\leq 1$. There is a further boost by the eccentricity, which will be presented in \S\ref{subsub:ecc}. 

In the single BH regime, $\UpTDE$ does not depend on $a$ any more (as the name implies). This indicates that $N_{\rm TDE}/N_{\rm tot}\propto a$ (e.g., $q=10^{-3}$ for $M_{\rm BH,1}=10^{7}\Msol$ in  Figure~\ref{fig:tde1} or $q=10^{-2}$ and $10^{-1}$ for $M_{\rm BH,1}=10^{8}\Msol$ in Figure~\ref{fig:tde2}). 

\subsubsection{Dependence on eccentricity}\label{subsub:ecc}

Like the $a$-dependence, $\UpTDE$ for SMBHBs in the primary BH-dominated binary regime depends on $e$ whereas its dependence is very weak in the other regimes.

In Figure~\ref{fig:ecc_dep}, we show $\UpTDE$ normalized by $\UpTDE$ at $e=0$ for binaries ($M_{\rm BH,1}=10^{7}\Msol$) with $q=10^{-3}$ (\textit{top-left}), $10^{-2}$ (\textit{top-right}), $10^{-1}$ (\textit{bottom-left}) and $1$ (\textit{bottom-right}), as a function of $e$. In the primary BH-dominated binary regime (i.e., $q\geq10^{-2}$, which applies to all the panels except for the \textit{top-left} panel), $\UpTDE$ increases with $e$ and is at most larger by a factor of $\simeq$3.5 than $\UpTDE(e=0)$. In the single BH regime ($q=10^{-3}$, \textit{top-left} panel), $\UpTDE$ remains nearly constant at order unity.  This is not surprising, provided the presence of the secondary BH is not important. 

It appears that $\partial(\Upsilon/\Upsilon(e=0))/\partial e$ is a function of $a/r_{\rm h}$ and $q$. For fixed $q$ and $e$, the slope decreases as $a/r_{\rm h}$ increases. This is most clearly seen for $q=10^{-2}$ (\textit{top-right}): $\partial(\Upsilon/\Upsilon(e=0))/\partial e\simeq 2.8$ for $a/r_{\rm h}=1$, $\simeq 2.1$ for $a/r_{\rm h}=10^{-1}$ and $\simeq1$ for $a/r_{\rm h}=10^{-2}$. On the other hand, as $q$ grows, the slope falls, which results in an almost constant value for $\Upsilon/\Upsilon(e=0)$ for $q=1$ within the range $0\leq e\leq 0.9$, independent of $a/r_{\rm h}$.
Although we only show $\Upsilon/\Upsilon(e=0)$ for binaries with $\pbhm=10^{7}\Msol$, the $e$-dependence is also quantitatively similar with that for other values of $\pbhm$ in this regime.

\subsection{Enhancement or suppression of direct captures by supermassive black hole binaries}

Similarly to $\UpTDE$, we define $\UpDC$ as,
\begin{align}\label{eq:rel_probaDC}
    \UpDC &=\left(\frac{N_{\rm DC}}{N_{\rm tot}}\right)\left(\frac{\mathcal{R}_{\rm DC}}{2a}\right)^{-1},
\end{align}
where $N_{\rm DC}$ is the number of DCs.
$\UpDC$ can be divided only into two regimes depending on $q$: the primary BH-dominated binary regime  ($q>10^{-3}$) and the effectively single BH regime with $M_{\rm BH}=\pbhm$ ($q\lesssim 10^{-3}$). $\UpDC$ in each regime has qualitatively the same dependence on $M_{\rm BH}$, $q$, $a$, $e$ as the same regime of $\UpTDE$, except that the cross-section is $\Sigma_{\rm DC}$. This is demonstrated in Figure~\ref{fig:dc1} where we show $\UpDC$ for circular SMBHBs with $\pbhm=10^{7}\Msol$ (\textit{left}) and $10^{8}\Msol$ (\textit{right}). Overall, the direct capture events can be enhanced by up to a factor of $\simeq 10$ (40) by circular (eccentric) SMBHBs relative to those by SMBHs.

The black hole mass dependence of $\UpDC$ is summarized in the \textit{right} panel of Figure~\ref{fig:parameter}.

\begin{figure}
	\centering
	\includegraphics[width=8.6cm]{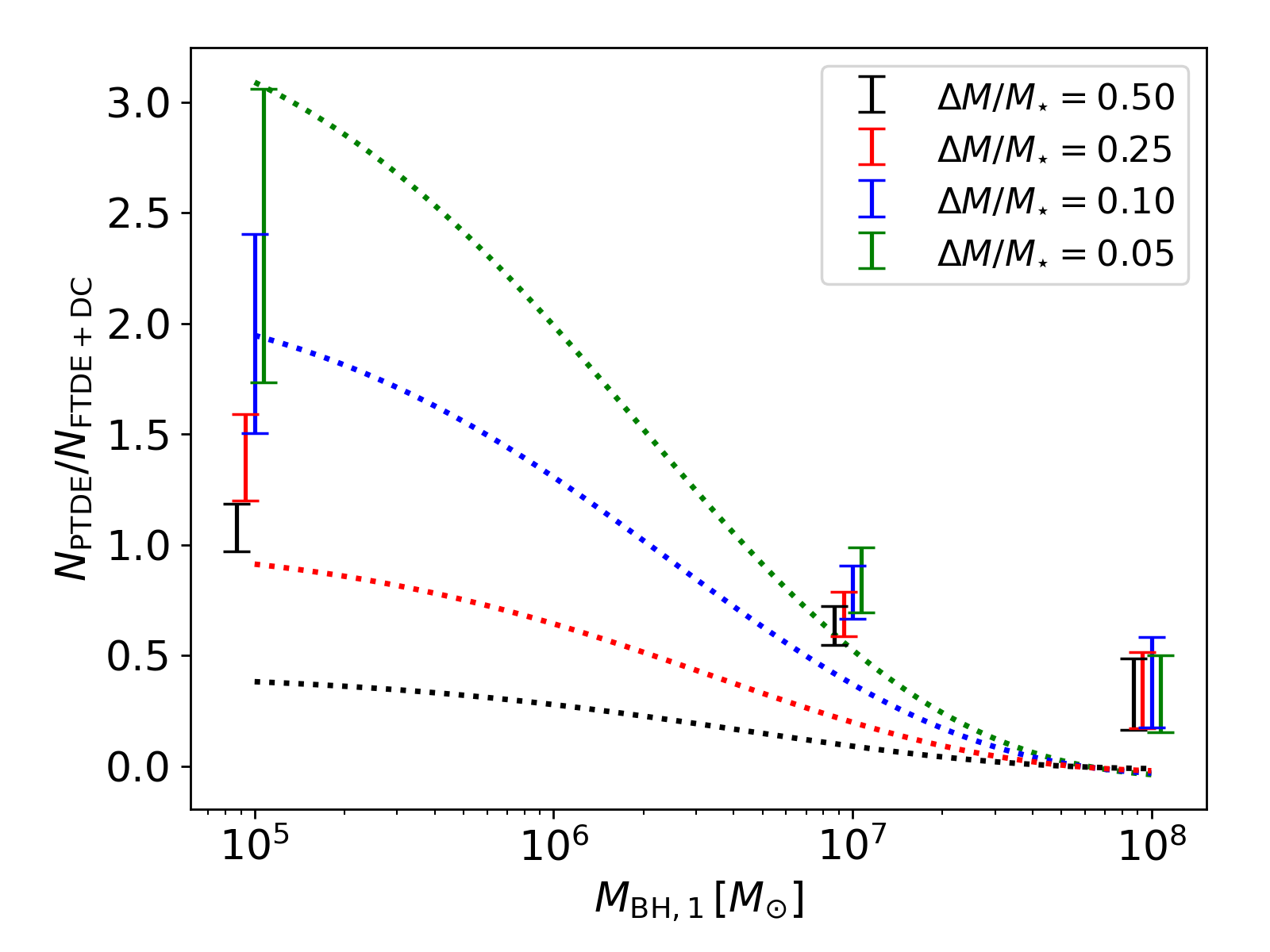}
	\caption{Ratio of the numbers of partial disruption events to that of star-destroying events (full disruptions and direct captures) as a function of the primary black hole mass. Each vertical bar show the entire variations of the number ratio for different $a$, $q$ and $e$ found in our simulations for a given fractional mass loss. The dotted lines depict the encounter cross section ratios between PTDEs and FTDEs+DCs by single SMBHs.}
	\label{fig:ptde2}
\end{figure}
\begin{figure*}
	\centering
	\includegraphics[width=8.6cm]{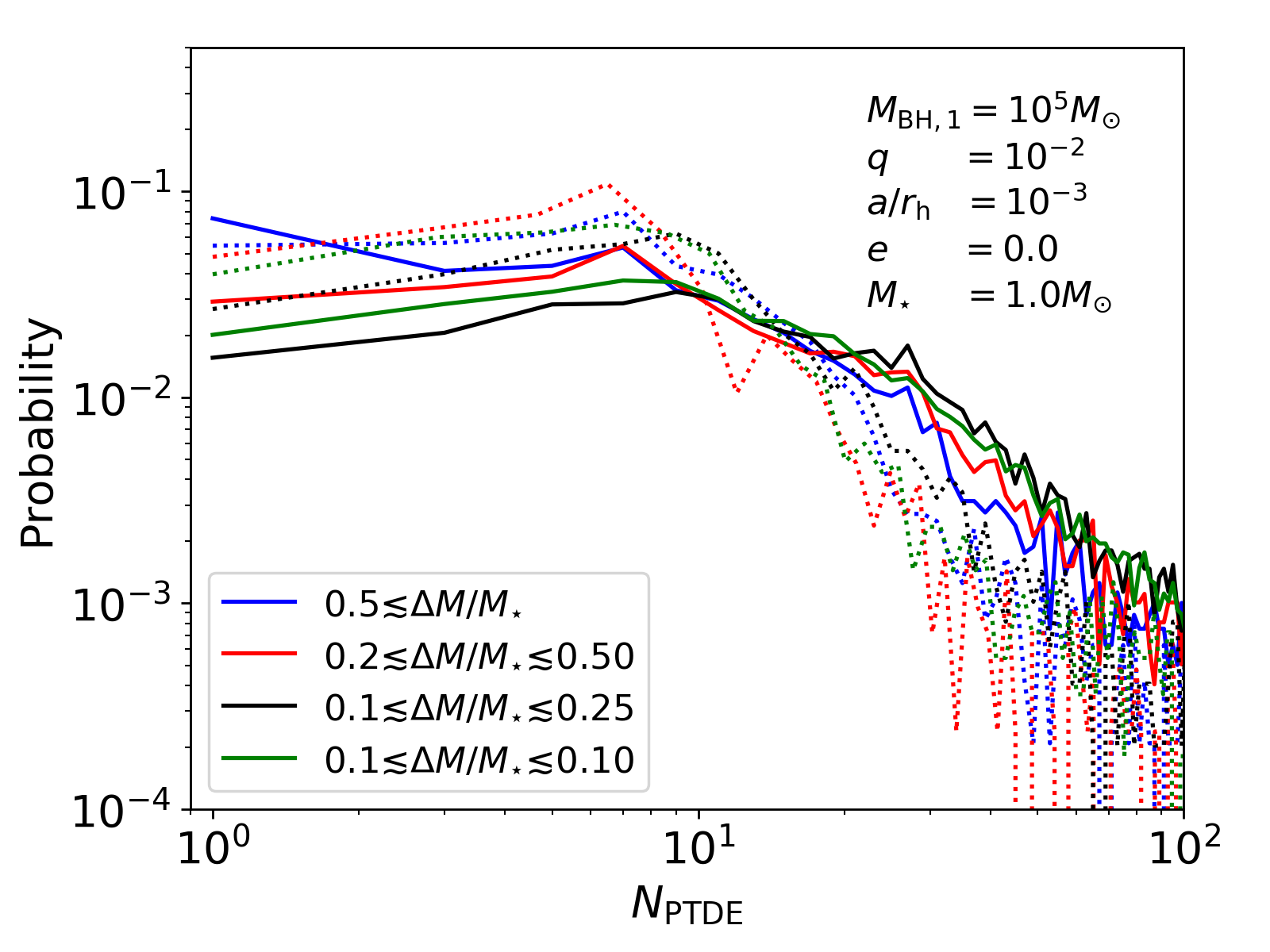}
	\includegraphics[width=8.6cm]{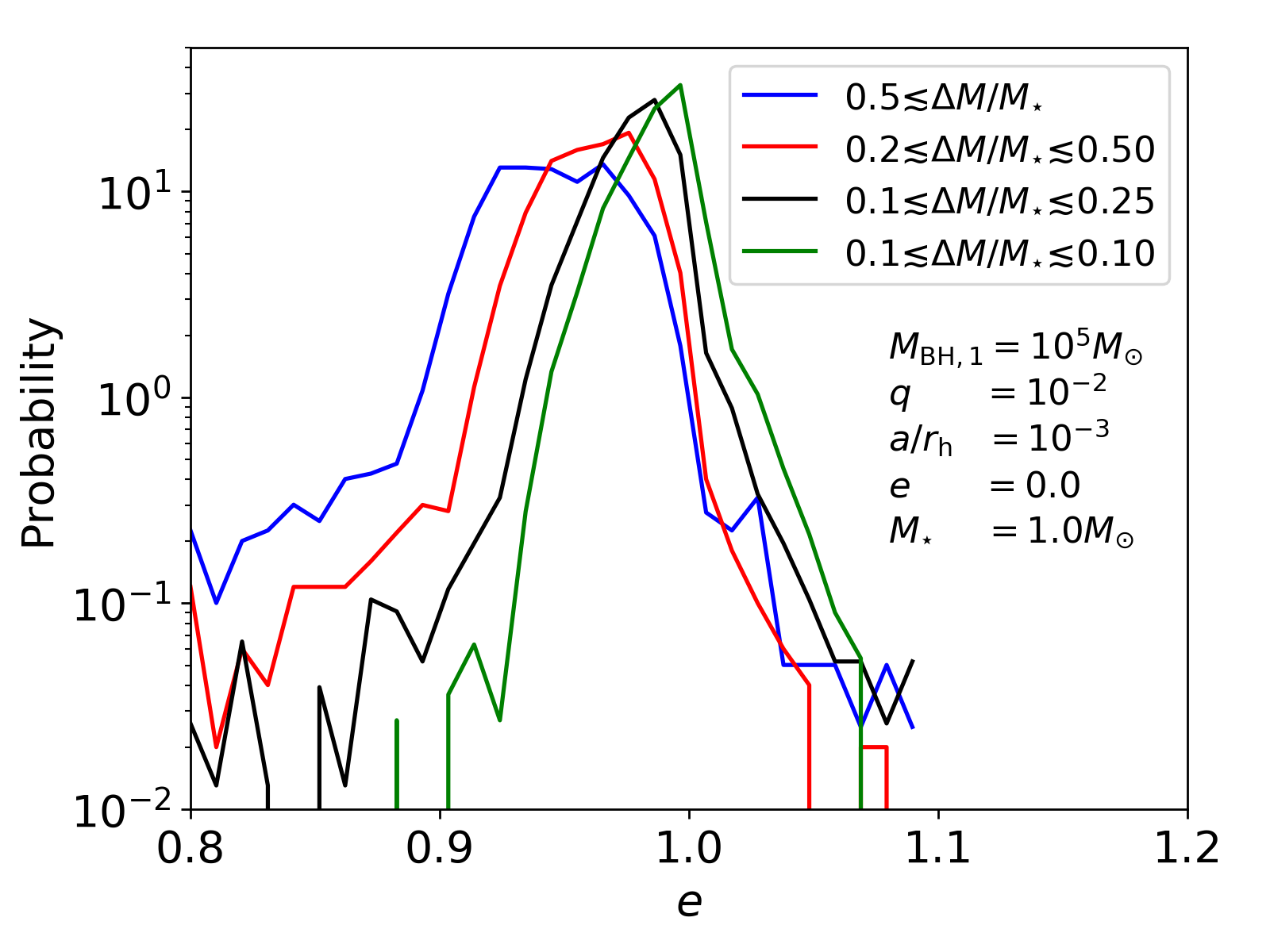}
	\caption{The probability for a star to undergo $N_{\rm PTDE}$ partial tidal disruptions (\textit{left}),
	 and the probability distribution for the eccentricity at the first close encounters (\textit{right}) around circular SMBHBs with $\pbhm=10^{5}\Msol$, $q=0.02$ and $a/r_{\rm h}=10^{-2}$ for a few different fractional mass losses $\Delta M/M_{\star}$. The eccentricity is measured at the first time-step after entering the PTDE regime.}
	\label{fig:ptde1}
\end{figure*}

\subsection{Impact of {relativistic effects}}\label{subsec:PN}

Comparing SMBHBs in the same regime, we find that relativistic effects tend to give higher values for $\UpTDE$ and $\UpDC$, and that the enhancement is greater for larger values of $\pbhm$. More specifically, $\UpTDE$ for $\pbhm<3\times 10^{7}\Msol$ is almost the same, independent of the PN terms, but the PN terms enhance $\UpTDE$ for $ 3\times 10^{7}\Msol\leq \pbhm<\widehat{M}_{\rm BH,1}$ by no more than a factor of $2-3$. We see similar trends in $\UpDC$, but the enhancement is somewhat greater: up to a factor of $5-10$ for $\pbhm>10^{7}\Msol$. We attribute the increase in the probability for larger $\pbhm$ to two main reasons: 1) stars have to get closer to more massive black holes to be disrupted, causing more frequent approaches and leading to greater energy loss via GWs on a per encounter basis and 2) relativistic effects increase the tidal radius (see Equation \ref{eq:RtBH}). 

\subsection{Fitting formula for $\UpTDE$}\label{subsec:fit}

To find a fit for $\UpTDE$, we performed a multivariate analysis for the cases with circular SMBHBs and the boosting effects associated with the binary eccentricity $e$, separately, using the {\sc SciPy} fitting tool {\sc curve\_fit} \citep{scipy}. Our fitting formula has the form
\begin{align}
    \UpTDE = \UpTDE(e=0) \UpTDE(e).
\end{align}
Note that we do not include the additional enhancement by the PN effects in our fitting formula. Nonetheless, because the PN terms increase $\UpTDE$ only for $\pbhm\simeq \widehat{M}_{\rm BH,1}$ by less than a factor of a few, we expect that our fitting formula can  capture the enhancement by SMBHBs for the parameter space where the majority of TDEs are expected to be created.
Our fitting formula for $\UpTDE(e=0)$ is,
\begin{align}\label{eq:upTDE_fit}
    \UpTDE(e=0) = \alpha~ [1.0 + (\pbhm/10^{6}\Msol)^{\beta}]|\log_{10}(a/r_{\rm h})+\gamma|^{\lambda} ~q^{\delta}~S^{\sigma}.
\end{align}
 where $\alpha$, $\beta$, $\lambda$, $\delta$ and $\sigma$ are all free parameters and $S$ are the term associated with the encounter cross section. 
For $\pbhm<\widehat{M}_{\rm BH,1}$,
\begin{align}
S&=\frac{\Sigma_{\rm TDE,1}+\Sigma_{\rm TDE,2}}{\Sigma_{\rm TDE+DC,1}+\Sigma_{\rm TDE+DC,2} },
\end{align}
and
\begin{align}
    \alpha &= 1.58,~ \beta = 0.0428,~ \gamma = 5.01, ~\lambda = 1.26,~ \delta = 0.263=4, ~\sigma = 0.111\nonumber\\ &\hspace{2.in}\text{for $q\geq 10^{-1}$},\\
    \alpha &= 42.4,~ \beta = 0.458,~ \gamma = -2.39, \lambda=-2.57,~ \delta = -0.0421,~ \sigma = 1.15\nonumber\\ &\hspace{2.in}\text{for $10^{-3}<q< 10^{-1}$},\\
    \alpha &= 1,~ \beta = 0, ~\delta = 0, ~\lambda = 0,~ \delta = 0,~ \sigma = 0\hspace{0.1in} \text{for $q\leq 10^{-3}$}.
\end{align}
For $\pbhm>\widehat{M}_{\rm BH,1}$ and $\sbhm<\widehat{M}_{\rm BH,2}$,
\begin{align}
    \alpha=14.4, ~\beta = 0,~ \gamma = 0,~ \lambda = 0, ~\delta = 0.657, ~S^\sigma = 1.
\end{align}
For $\pbhm, \sbhm>\widehat{M}_{\rm BH,2}$, 
\begin{align}
    \UpTDE(e=0)=0.
\end{align}
The predicted values from the fitting formula are shown in Figures~\ref{fig:tde1} and \ref{fig:tde2} using dashed lines.

The boosting effects $\UpTDE(e)$ due to the eccentricity $e$ are described by, for $\pbhm<\widehat{M}_{\rm BH,1}$
\begin{align}\label{eq:ecc_boost}
  \UpTDE(e)=
    \begin{cases}
    \frac{10.6}{|\log_{10}(a/r_{\rm h})|^{2.8}+4.18} \left(\frac{|\log_{10}q|}{2}\right)^{1.48} e +1\hspace{0.1in}\text{for $q>10^{-3}$} ,\\
  \frac{0.433}{|\log_{10}(a/r_{\rm h})|^{16.36}+0.623} \left(\frac{|\log_{10}q|}{2}\right)^{-0.377} e +1 \hspace{0.1in}\text{for $q\leq10^{-3}$},
    \end{cases}
\end{align}
and for $\pbhm>\widehat{M}_{\rm BH,1}$,
\begin{align}
    \UpTDE(e)= 1,
\end{align}
which are shown in Figure~\ref{fig:ecc_dep} using dashed lines.

\subsection{Partial disruption events}\label{sec:ptde}

When a star orbits at a distance slightly greater than the tidal radius, the tidal forces are not sufficiently large to completely disrupt the star. As a result, the star survives after it loses some fraction of its mass. This event is called a partial tidal disruption event (PTDE). PTDEs can cause several impacts on the evolution of the star. The orbits can be significantly different before and after the pericenter passage as a result of the orbital energy transfer due to tidal excitation and asymmetric mass loss \citep{Manukian+2013,Ryu+2020c}. Furthermore, upon partial mass loss, the remnant is out of thermal equilibrium and can have high spin values due to tidal torques, even when the star initially does not rotate \citep{Ryu+2020c}. High entropy and rapid rotation puff up the star, which makes it more or less subject to full and partial disruptions depending on the spin axis relative to the orbital axis at the next pericenter passage \citep{Golightly+2013}. 

For single SMBHs, partial disruptions are more frequent than full disruptions \citep{StoneMetzger2016,Krolik+2020}. The fact that the probabilities for FTDEs by SMBHBs are proportional to their cross sections and enhanced compared to those by single SMBHs naturally suggests a similar relation between the probability for PTDEs and their cross sections, as well as an enhancement due to the existence of the secondary BH. Furthermore, the chaotic and secular three-body interactions can facilitate multiple PTDEs on a timescale shorter than the lifetime of the star. Considering their impact on the stars, PTDEs, if they happen, can potentially change the the overall final fate of the star. In this section, we investigate the event probability of PTDEs and the orbital properties of the stars at such close encounters.

For this analysis, we make use of the simulations for $\pbhm=10^{5}\Msol$, $10^{7}\Msol$ and $10^{8}\Msol$ with the PN-terms switched on, in which every pericenter distance smaller than $2 r_{\rm t}$ is recorded before the simulation is terminated. Although we did not simulate PTDEs in our simulations, we can make a rough estimate of the mass loss at each pericenter passage. To achieve this, we first find a fitting formula for the relations between the mass loss and the pericenter distance for $\mstar=1\Msol$ from \citet{Ryu+2020d} (their Figure 4), 
\begin{align}
    \Delta M/M&=(r_{\rm p}/\mathcal{R}_{\rm t})^{-\alpha},
\end{align}
where $\log_{10}\alpha = 3.15\times10^{-8} [\log_{10} (M_{\rm BH}/M_{\odot})]^{8.42} + 0.3$. By inverting this formula, we can estimate the pericenter distance for a given fractional mass loss.  Note that one caveat of this analysis is that because the mass loss curves from \citet{Ryu+2020d} are found for events where stars are initially on parabolic orbits, the mass loss could be different for highly eccentric ($e\simeq0.9$) or slightly hyperbolic ($e\lesssim 1.05$) orbits, but not by a significant amount \footnote{To confirm this conjecture, we performed a few fully relativistic hydrodynamical simulations for partial disruptions of a middle-age main-sequence $1\Msol$ star on a very eccentric orbit ($e=0.9$). We find that the remnant mass from the star on an initially eccentric orbit is roughy $\simeq 10\%$ smaller than that from the star on an initially parabolic orbit at the same pericenter distance.}. However, we only use this mass fraction estimate to make qualitative arguments for possible high frequencies of multiple severe partial disruption events.

We find that the number of PTDEs can be larger by up to a factor of $3$ than the number of FTDEs for given SMBHB parameters. In Figure~\ref{fig:ptde2}, we show the ratio of the number of PTDEs $N_{\rm PTDE}$ to that of the star-removing events  $N_{\rm FTDE+DC}$ in our scattering experiments, as a function of $\pbhm$. The vertical lines show the range of the event fraction across the ranges of $q$ and $a$ considered. For better readability, we shift the vertical lines horizontally by $\Delta M_{\rm BH,1}=10^{0.1}$. 
For comparison, we depict the cross section ratio $(\Sigma_{\rm PTDE}-\Sigma_{\rm FTDE+DC})/ \Sigma_{\rm FTDE+DC}$ for $\pbhm$ using dotted lines. For $M_{\rm BH}\leq 10^{7}\Msol$, the event fraction is generally larger for smaller $\Delta M / M_{\star}$, which is exactly the same trend with the cross section ratio, as expected. 

These PTDEs generally occur in more than 50-90\% of all of the FTDE-terminated events by $\pbhm=10^{5}\Msol$.
The percentile decreases with $\pbhm$: $40-70\%$ for $\pbhm=10^{7}\Msol$ and $35-50\%$  for $\pbhm=10^{8}\Msol$. This fraction seems to increase with $a/r_{\rm h}$, while it is weakly dependent on $q$ and $e$.

We also find that severe PTDEs can happen a few times before our termination events. The \textit{left} panel of Figure~\ref{fig:ptde1} shows the probability density function of the number of PTDEs for $1\Msol$ stars by circular SMBHBs with $\pbhm=10^{5}\Msol$, $q=10^{-2}$ and $a/r_{\rm h}=10^{-2}$. The distribution is more or less flat for $N_{\rm PTDE}\leq 10$ with a peak at $\simeq 1$ for $\Delta M /M_{\star}\gtrsim 0.5$, and at $10$ for  $\Delta M /M_{\star}\lesssim 0.5$. For the simulations that end with FTDEs (dotted lines), the events for $\Delta M /M_{\star}\geq0.5$ occur more frequently (i.e., they peak at $\simeq 10$). Although we show this for one case, we find similar results for the other cases. We will discuss its implication in Section~\ref{subsec:disc_ptde}. 

The orbits for partial disruptions by SMBHBs are not always parabolic. The orbits become more eccentric for closer encounters. The \textit{right} panel of Figure~\ref{fig:ptde1} shows the probability distribution function for $e$ at the first close encounter that can yield non-zero mass loss for the same SMBHB shown in the \textit{left} panel.
The peak of the distribution shifts to lower $e$ as the encounter becomes closer. We find similar distributions for all other cases, except for $a/r_{\rm h}=10^{-4}$ for which the distribution becomes flatter for $e\lesssim 1.1$.

\section{Discussion}\label{sec:discussion}

In this section, we discuss the significance of our results for observations of tidal disruption events, along with possible improvements to our work that can be implemented in future studies.

\subsection{Partial disruption events}\label{subsec:disc_ptde}

We find that before stars are ejected, fully disrupted or directly captured by SMBHBs, there are multiple close encounters that could lead to severe partial disruptions in our simulations. The multiple encounters found in our simulations do not directly imply that the stars would go through multiple partial disruptions because, as mentioned in \S\ref{sec:ptde}, the orbits of the remnants could be significantly different from the initial stellar orbit. However, it is quite clear that the stars would go through at least one partial disruption when the star enters the partial disruption zone for the first time. Furthermore, multiple occurrences that we find can imply that, as long as the stars remain bound to the SMBHBs after partial disruptions, there can be a high probability that stars will enter the partial disruption zone multiple times. 

This possibility is shown indirectly in \citet{Chen+2008,Chen+2009} as they present the close encounter probability in terms of the cross-section. However, they did not remark on the potential significance of partial disruptions. Later, \citet{Coughlin+2017} reported multiple close encounters in their simulations in which they considered interactions between a $1\Msol$ star and an SMBHB with $\pbhm=10^{6}\Msol$ and $0.1\leq q\leq 1$ and $a/r_{\rm h}\simeq 4\times 10^{-4}$, and mentioned the possibility of partial disruption events. In this paper, we confirm that these potential partial disruption events are in fact very frequent over a wide range of parameter space. We remark that in the previous works they assume that stars are fully disrupted when the pericenter distance is smaller than $r_{\rm t}$. But realistic middle-aged radiative stars ($M_{\star}\geq 0.7\Msol$) are not fully disrupted at $r_{\rm p}=r_{\rm t}$, but are instead only partially disrupted \citep{Golightly+2013,Law-Smith+2019,Ryu+2020a,Nixon+2021}. \citet{Ryu+2020c} showed that realistic $1\Msol$ stars would lose only 10\% of their initial mass at $r_{\rm p}\simeq r_{\rm t}$ (see their Figure 4) in partial disruptions by a $10^{6}\Msol$ BH when relativistic effects are taken into account. Therefore, the estimates for the disruption probability in previous studies effectively correspond to the probabilities for weak partial disruption events. 

Multiple partial disruptions can have important observational implications because multiple partial disruptions could result in light curves that can not be explained by full disruptions. Also, the true FTDE rate by SMBHBs could be different from what is predicted based solely on the frequency of FTDEs in other numerical simulations. As an extreme case, if it is frequent that PTDEs occur multiple times before a star is fully disrupted, full disruptions are rare and the light curves associated with such events can reveal quasi-periodic behavior. Therefore, the theoretical prediction of the SMBHB-driven TDE rate should be properly corrected to account for this effect and be representative of the true rate.

\subsection{Caveats and future improvements}\label{subsec:caveats}

We estimate the probability of FTDEs by SMBHBs relative to that for single SMBHs assuming the tidal radii for typical TDEs by single SMBHs. For TDEs by SMBHBs, the star can undergo multiple encounters prior to the one yielding FTDEs, which can affect the tidal radius due to altered internal structure \citep{Ryu+2020a,Ryu+2020c} and stellar spin \citep{Golightly+2013}\footnote{The tidal radius can be different depending on the eccentricity of the star's orbit \citep{Cufari+2022}. }. Therefore, to more accurately measure the TDE probability, it is necessary to consider partial disruption events and their impact on the tidal radius. However, the tidal radius and the adjusted orbits of partially disrupted stars are not fully understood. In particular, if the time between close encounters is shorter than the time scale for spin angular momentum dissipation, the correction of the tidal radius should be taken into account. In addition, because the spin axis of remnants are not always aligned with the orbital plane at the next encounter because of the chaotic nature of three-body interactions, the dependence of the remnant properties on stellar spins should be investigated further. This will be the focus of a future study.

We do not consider perturbations from surrounding stars and remnants in the vicinity of the SMBHBs.  However, \citet{reinoso22} recently showed using N-body simulations that in dense environments perturbations in angular momentum-space operate on very short timescales, and can cause significant changes in eccentricity.  This could translate into significantly perturbing stars that survive the first partial disruption event, altering the orbital properties more than accounted for in this study.  Thus, the results in this paper should be confirmed using actual N-body simulations to properly model the surrounding stellar distribution and its effects on the rates of FTDEs and PTDEs.  \citet{vergara21} recently performed such an experiment.  The authors account for cluster rotation, which is an important parameter that should be considered in more detail in future studies.  For example, as suggested by \citet{webb19}, the cluster rotation could reduce the relative velocity between a black hole in the central regions of the cluster and orbiting stars, potentially enhancing the rates of PTDEs and FTDEs.

In this study, we did not attempt to perform detailed calculations of the event rate which require a careful modeling of the rate at which the stellar flux enters the parameter space for full disruptions (so called tidal disruption loss cone) in different types of galaxies, stellar mass function, and black hole mass function, evolutionary stage of SMBHBs \citep{Chen+2011} and electromagnetic emission from the events (e.g., modulation of light curves \citet{Coughlin+2017}). The TDE rates by SMBHBs estimated in previous work suggest that the TDE rate is very dependent on the assumptions for the loss cone refilling \citep{Chen+2008, Darbha+2018}. As we stressed above, potentially frequent partial tidal disruption events make the rate calculations more complicated. We will investigate the rate which takes into account partial tidal disruptions in our future work.

\section{Conclusion}\label{sec:concluson}

We performed a large number of highly-accurate three-body scattering experiments to investigate full disruption and direct capture events by relatively compact supermassive black hole binaries. We examined these events for wide ranges of key parameters ($10^{5}\Msol\leq \pbhm\leq 10^{7}\Msol$, $10^{-3}\leq q\leq 1$, $10^{-4}\leq a/r_{\rm h}\leq 1$, $0\leq e \leq 0.9$ and $0.3\Msol \leq \mstar \leq 3\Msol$), which is a significant extension of the parameter space explored in previous work. In addition, we studied partial disruption events that can lead to a fractional mass loss and orbital element changes prior to the star-removing events and discussed their implications. 

We can summarize our results as follows,

\begin{itemize}
    \item The encounter probabilities of FTDEs and DCs by hard SMBHBs are well-described by the encounter cross section (\S \ref{sec:crosssection}), which is proportional to $\mathcal{R}_{\rm t}$ and $\mathcal{R}_{\rm DC}$, respectively. In particular, the $M_{\star}$-dependence of the encounter probability and event fraction is fully incorporated in the encounter cross section (see Figure~\ref{fig:eventfraction_mstar}).
    
    \item FTDEs by SMBHBs can be divided into four different regimes depending on the primary mass and mass ratio. For $\pbhm<\widehat{M}_{\rm BH,1}$ and $q>10^{-3}$, FTDEs can be enhanced by up to a factor of $30-40$ at a rate that depends on $a$, $q$ and $e$. However, for $\pbhm<\widehat{M}_{\rm BH,1}$ and $q\lesssim 10^{-3}$, the existence of the secondary BH does not enhance FTDEs, which makes the SMBHBs effectively act like a single BH with $M_{\rm BH}=\pbhm$. When $\pbhm>\widehat{M}_{\rm BH,1}$, only the secondary BH can disrupt stars. For this case, FTDEs are slightly enhanced ($q>10^{-3}$, or the single BH regime with $M_{\rm BH}=\sbhm$) or significantly suppressed ($q\leq10^{-3}$, or the no BH regime). See the \textit{left} panel of Figure~\ref{fig:parameter}).

    \item  Unlike FTDEs, there is no black hole mass constraint for DCs. This results in dividing DCs by SMBHBs into two regimes, depending on the mass ratio. For $q>10^{-3}$, the existence of the secondary BH can significantly enhance the DC probability at a rate that depends on $a$, $q$ and $e$. On the other hand, for $q\lesssim 10^{-3}$, as with FTDEs, SMBHBs act like a single SMBH and there is no significant enhancement in DCs. See the \textit{right} panel of Figure~\ref{fig:parameter}.

    \item  We provide a fitting formula for FTDE enhancement by SMBHBs that works for a wide range of parameter space (see \S\ref{subsec:fit}).

    \item Relativistic effects tend to increase $\UpTDE$ and $\UpDC$ by an amount that is larger for larger $\pbhm$. $\UpTDE$ ($\UpDC$) can be enhanced by no more than a factor of $2-3$ ($5-10$) when relativistic effects are not included (\S\ref{subsec:PN}).  

    \item  The partial disruption events can occur more frequently than FTDEs by a factor of three (Figure~\ref{fig:ptde1}). In addition, we find that stars orbit with $0.9\lesssim e\lesssim1$ close enough to lose a large fraction of their mass multiple times before our termination events (FTDEs and DCs) (Figure~\ref{fig:ptde2}). Because such partial disruption events can induce stellar spins and mass loss, which in turn changes the tidal radius and orbital parameters, they can significantly affect the overall full disruption event rate and the shape of the light curves.

\end{itemize}

\section*{Acknowledgements} 
This research project was conducted using computational resources (and/or scientific computating services) at the Max-Planck Computing \& Data Facility. AAT received support from JSPS KAKENHI Grant Numbers 17H06360, 19K03907 and 21K13914.
N.W.C.L. acknowledges the support of a Fondecyt Iniciación grant 11180005, the financial support from Millenium Nucleus NCN19-058 (TITANs) and the BASAL Centro de Excelencia en Astrofisica y Tecnologias Afines (CATA) grant CATA AFB170002 along with ANID BASAL projects ACE210002 and FB210003.

\section*{Data Availability}
Any data used in this analysis are available on reasonable request from the first author.

\bibliographystyle{mnras}
%\bibliography{biblio.bib} % if your bibtex file is called example.bib

\appendix

\section{Comparison with $R_{\rm TDE}=\lowercase{r_{\rm t}}$}

\begin{figure*}
	\centering
	\includegraphics[width=8.6cm]{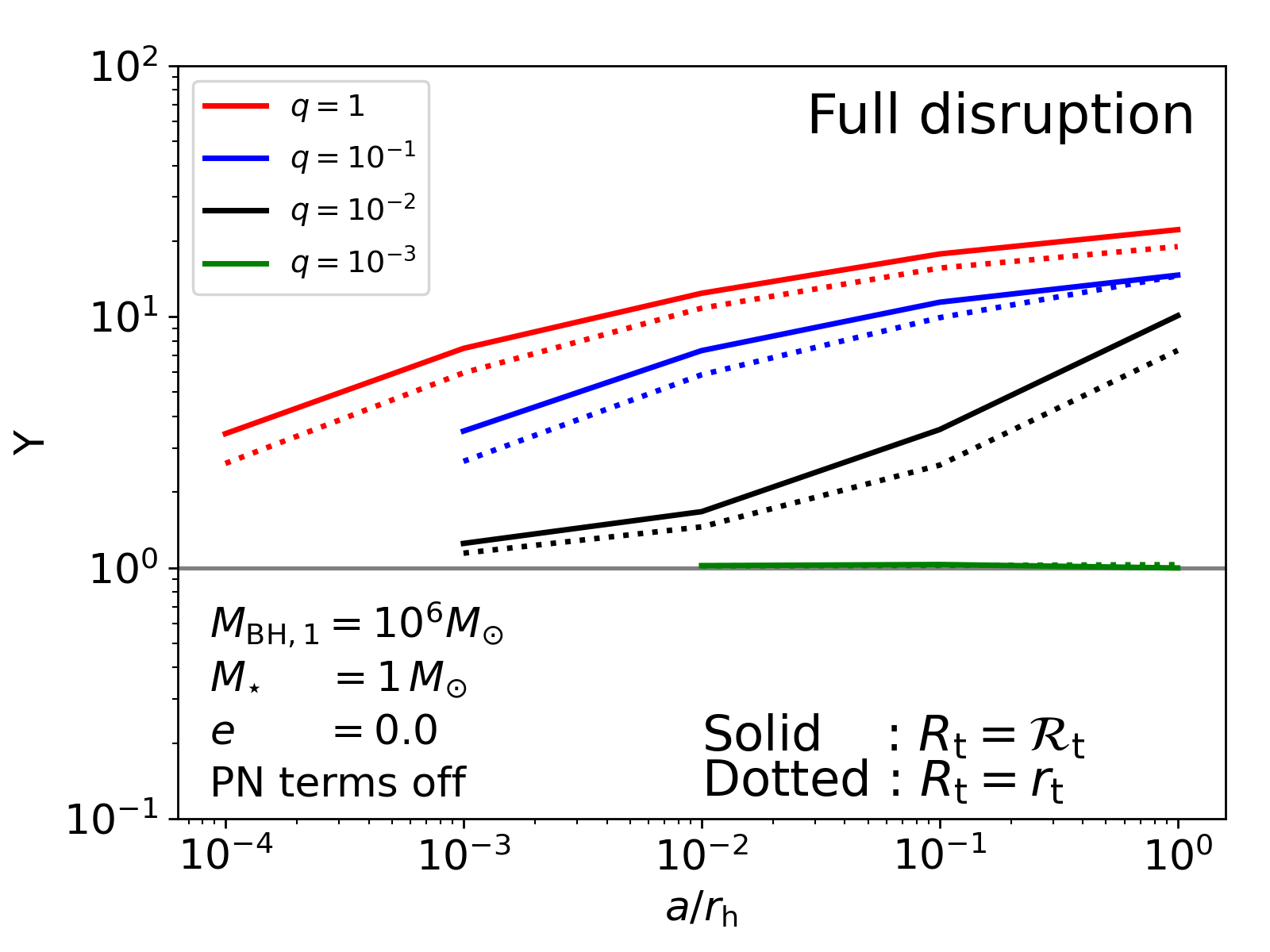}
	\includegraphics[width=8.6cm]{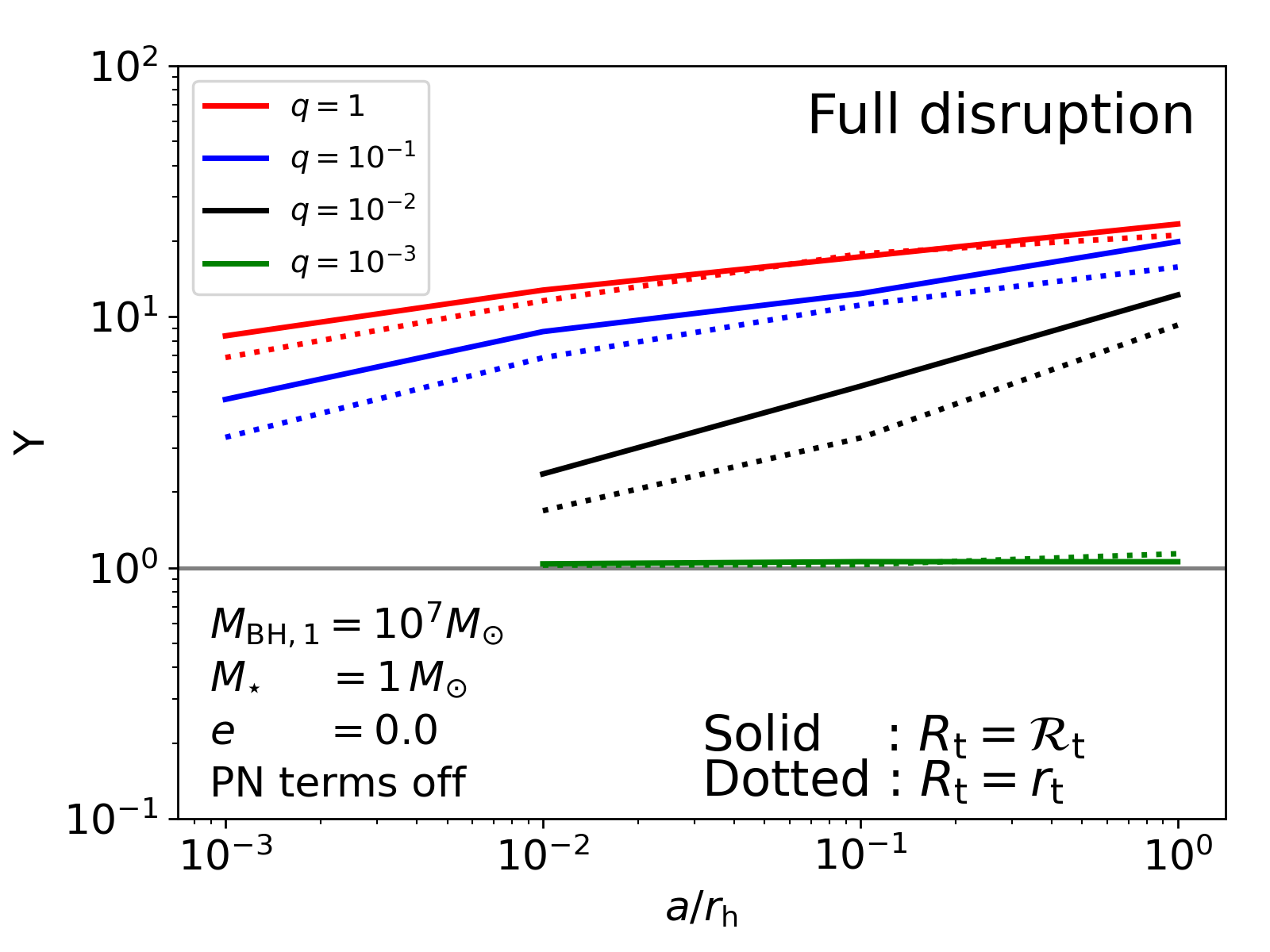}
	\caption{The relative probability $\UpTDE$ with $\mathcal{R}_{\rm TDE}=r_{\rm t}$ for $\pbhm=10^{6}\Msol$ (\textit{left}) and $10^{7}\Msol$ (\textit{right}) when the PN terms are off. The solid lines indicate $\UpTDE$ with $R_{\rm TDE}=\mathcal{R}_{\rm t}$ and dotted lines that with $R_{\rm TDE}=r_{\rm t}$. }
	\label{fig:appn1}
\end{figure*}

To compare if the relative probability $\UpTDE$ is robust against a different value of $R_{\rm TDE}$, we performed extra simulations for circular SMBHBs with $\pbhm=10^{6}\Msol$ and $10^{7}\Msol$ and $10^{-3}\leq q\leq 1$ and $10^{-4}\leq a/r_{\rm h}\leq 1$ where $\Psi=1$ (or $R_{\rm TDE}=r_{\rm t}$). The results are presented in Figure~\ref{fig:appn1} where $\UpTDE$ are calculated using $r_{\rm t}$ in place of $\mathcal{R}_{\rm t}$ in Equation~\ref{eq:rel_proba}. The solid lines indicate $\UpTDE$ with $R_{\rm TDE}=\mathcal{R}_{\rm t}$ and the dotted lines that with $R_{\rm t}=r_{\rm t}$. We find that they are in good agreement with the difference in $\UpTDE$ smaller than $\simeq 20\%$. Although we only show $\UpTDE$, we also find a similar level of consistency in $\UpDC$ for both cases.

\end{document}